\documentclass[10pt,twocolumn,letterpaper]{article}

\usepackage{cvm}
\usepackage{times}
\usepackage{epsfig}
\usepackage{graphicx}
\usepackage{amsmath}
\usepackage{amssymb}

\usepackage{subfigure}
\usepackage{multirow}
\usepackage{amsmath}
\usepackage{amssymb}
\usepackage{mathrsfs}
\usepackage{xcolor}
\usepackage{authblk}

\usepackage[pagebackref=true,breaklinks=true,letterpaper=true,colorlinks,bookmarks=false]{hyperref}

\cvmfinalcopy
\def\cvmPaperID{****} % *** Enter the cvm Paper ID here

\newcommand{\yyj}[1]{{\color{black}#1}}

\ifcvmfinal\fi
\begin{document}

%%%%%%%%% TITLE
\title{Mesh Variational Autoencoders with Edge Contraction Pooling}

%\author{First Author\\
%Institution1\\
%Institution1 address\\
%{\tt\small firstauthor@i1.org}
%% For a paper whose authors are all at the same institution,
%% omit the following lines up until the closing ``}''.
%% Additional authors and addresses can be added with ``\and'',
%% just like the second author.
%% To save space, use either the email address or home page, not both
%\and
%Second Author\\
%Institution2\\
%First line of institution2 address\\
%{\small\url{http://www.author.org/~second}}
%}

\author[a,b]{Yu-Jie Yuan}
\author[c]{Yu-Kun Lai}
\author[a,b]{Jie Yang}
\author[d]{Hongbo Fu}
\author[a*]{Lin Gao}
\affil[a]{Beijing Key Laboratory of Mobile Computing and Pervasive Device, Institute of Computing Technology, CAS}
\affil[b]{University of Chinese Academy of Sciences}
\affil[c]{School of Computer Science and Informatics, Cardiff University, UK}
\affil[d]{City University of Hong Kong}

\renewcommand*{\Affilfont}{\small\it} % 修改机构名称的字体与大小
\renewcommand\Authands{ and } % 去掉 and 前的逗号

\maketitle
\if TT\insert\footins{\noindent\footnotesize{*Corresponding Author: gaolin@ict.ac.cn (Lin Gao)}}\fi

\begin{abstract}
	3D shape analysis is an important research topic in computer vision and graphics. While existing methods have generalized image-based deep learning to meshes using graph-based convolutions, the lack of an effective pooling operation restricts the learning capability of their networks. In this paper, we propose a novel pooling operation for mesh datasets with the same connectivity but different geometry, by building a mesh hierarchy using mesh simplification. For this purpose, we develop a modified mesh simplification method to avoid generating highly irregularly sized triangles. Our pooling operation effectively encodes the correspondence between coarser and finer meshes in the hierarchy. We then present a variational auto-encoder structure with the edge contraction pooling and graph-based convolutions, to explore probability latent spaces of 3D surfaces. Our network requires far fewer parameters than the original mesh VAE and thus can handle denser models thanks to our new pooling operation and convolutional kernels. Our evaluation also shows that our method has better generalization ability and is more reliable in various applications, including shape generation, shape interpolation and shape embedding.
	%Our framework requires  thanks to convolutional kernels \hongbo{you mean the original mesh VAE doesn't use convolutional kernels?} and pooling,
	%and can process denser models. Extensive experiments show that our network has better generalization ability thanks to our new pooling operation. % , and the new pooling operation effectively improves the generalization ability. 
	%Moreover, our framework is able to produce more reliable results in various applications, including shape generation, shape interpolation and shape embedding.
\end{abstract}

%%%%%%%%% BODY TEXT
\section{Introduction}

In recent years, 3D shape datasets have been increasingly available on the Internet. Consequently, data-driven 3D shape analysis has been an active research topic in computer vision and graphics. Apart from traditional data-driven works such as \cite{CGF:CGF12991}, recent works attempted to generalize deep neural networks from images to 3D shapes such as \cite{tan2018aaai,meshVAE,Litany_2018_CVPR} for triangular meshes, \cite{qi2016pointnet} for point clouds, \cite{3dgan,voxnet} for voxel data, and so on. In this paper, we concentrate on deep neural networks for triangular meshes. Unlike images, 3D meshes have complex and irregular connectivity. Most existing works tend to keep mesh connectivity unchanged from layer to layer, thus losing the capability of increased receptive fields when pooling operations are applied.

As a generative network, the Variational Auto-Encoder (VAE) \cite{kingma2013auto} has been widely used in various kinds of generation tasks, including generation, interpolation and exploration on triangular meshes \cite{meshVAE}.
The original MeshVAE \cite{meshVAE} uses a \emph{fully connected} network that requires a huge number of parameters and its generalization ability is often weak. Although the fully connected layers allow the changes of mesh connectivity, due to irregular  changes, they cannot be followed by convolutional operations. Litany et al.~\cite{Litany_2018_CVPR} use the VAE structure with graph convolutions, aiming to deal with model completion. Gao et al. \cite{gao2018automatic} include a spatial convolutional mesh VAE in their pipeline. However, these two kinds of convolution operations cannot change the connectivity of the mesh. The work~\cite{ranjan2018generating} introduces sampling operations in convolutional neural networks (CNNs) on meshes, but their sampling strategy does not aggregate all the local neighborhood information when reducing the number of vertices. Therefore, in order to deal with denser models and enhance the generalization ability of the network, it is necessary to design a pooling operation for meshes similar to the pooling for images to reduce the number of network parameters. Moreover, it is desired that the defined pooling operation can support further convolutions and allow recovery of the original resolution through a relevant de-pooling operation.

In this paper we propose a VAE architecture with newly defined pooling operations. Our method uses mesh simplification to form a mesh hierarchy with different levels of details, and achieves effective pooling by keeping track of the mapping between coarser and finer meshes. To avoid generating highly irregular triangles during mesh simplification, we introduce a modified mesh simplification approach based on the classical mesh simplification algorithm \cite{garland1997surface}. The input to our network is a vertex-based deformation feature representation~\cite{gao2017sparse}, which unlike 3D coordinates, encodes deformations using deformation gradients defined on vertices. \yyj{Our framework use a collection of 3D shapes with the same connectivity to train the network. Meshes with consistent connectivity have been commonly used for various applications such as data-driven deformation~\cite{sumner2004deformation}, deformation transfer~\cite{gao2018automatic,SDT}, human shape generation \cite{meshVAE} and shape completion~\cite{Litany_2018_CVPR}. Such meshes can be easily obtained through consistent remeshing.} 
Also, we adopt graph convolutions~\cite{defferrard2016convolutional} in our network. In all, our network follows a VAE architecture where pooling operations and graph convolutions are applied. As we will show later, our  network not only has better generalization capabilities but also can handle much higher resolution meshes, benefiting various applications, such as shape generation, interpolation and embedding.

\section{Related Work}

\noindent\textbf{Deep Learning for 3D Shapes.} Deep learning on 3D shapes has received increasing attention. {From the Euclidean domain, Boscaini et al.~\cite{boscaini2016learning,boscaini2016anisotropic} generalize CNNs to the non-Euclidean domain, which is useful for 3D shape analysis such as establishing correspondences.}
Bronstein et al.~\cite{bronstein2017geometric} give an overview of utilizing CNNs on non-Euclidean domains, including graphs and meshes. 
Masci et al.~\cite{masci2015geodesic} proposed the first mesh convolutional operations by applying filters to local patches represented in geodesic polar coordinates. 
Sinha et al.~\cite{sinha2016deep} convert 3D shapes to geometry images to obtain a Euclidean parameterization, on which standard CNNs can be applied. 
Maron et al.~\cite{maron2017convolutional} parameterize a surface to a planar flat-torus to define a natural convolution operator for CNNs on surfaces. 
Wang et al.~\cite{Wang-2017-ocnn,Wang-2018-aocnn} proposed octree-based convolutions for 3D shape analysis.
Unlike local patches, geometry images, planar flat-torus, or octree structure, our work performs convolutional operations using vertex features~\cite{gao2017sparse} as inputs.
%\hongbo{similar to the comments I made here, instead of listing the existing works, you should put our work in the context of the existing ones by discussing how our work is related to or different from the existing ones. It's a common practice to have such a discussion at the end of each paragraph in the related work section.}
%\hongbo{The above works and (some of) the following works take different inputs (a single model versus deforming meshes)? This should be made clear and try not to mix such discussions without telling the key differences/purposes.}

To analyze meshes with the same connectivity but different geometry, the work~\cite{meshVAE} first introduced the VAE architecture to 3D mesh data, and demonstrates its usefulness using various applications. Tan et al.~\cite{tan2018aaai} use a convolutional auto-encoder to extract localized deformation components from mesh datasets with large-scale deformations. Gao et al.~\cite{gao2018automatic} proposed a network which combines convolutional mesh VAE with CycleGAN~\cite{zhu2017} for automatic unpaired shape deformation transfer. The works of~\cite{tan2018aaai,gao2018automatic} apply convolutional operations to meshes in the spatial domain, while the works of~\cite{defferrard2016convolutional,henaff2015deep} extend CNNs to irregular graphs by construction in the spectral domain, and show superior performance when compared with spatial convolutions. Following ~\cite{defferrard2016convolutional,yi2017syncspeccnn}, our work also performs convolutional operations in the spectral domain. 

While pooling operations have been widely used in the deep networks for image processing, existing mesh-based VAE methods either do not support pooling~\cite{meshVAE,gao2018automatic}, or use a simple sampling process~\cite{ranjan2018generating}, which is not able to aggregate all the local neighborhood information. In fact, the sampling approach in~\cite{ranjan2018generating}, while being also based on a simplification algorithm, directly drops the vertices, and uses the barycentric coordinates in a triangle to recover the lost vertices by interpolation. In contrast, our pooling operations can aggregate local information by recording the simplification procedure, and support direct reverse of the pooling operation to effectively achieve a de-pooling operation.

\yyj{\noindent\textbf{Uniform Sampling or Pooling Methods.} Based on point cloud, Pointnet++ \cite{pointnet++} has proposed a uniform sampling method for point cloud based neural networks. Using the same idea, TextureNet \cite{texturenet} also conducts uniform sampling on the vertices of mesh. This kind of sampling method destroys the connection between vertices, turning mesh data into point cloud and cannot support further graph convolutions. Simplification methods can build mesh hierarchy, so can help us build mesh pooling operation. However, most simplification methods, such as \cite{garland1997surface}, are shape-preserving, not uniform. Remeshing operation, such as \cite{remesh}, can build uniform simplified meshes, but loss the correspondence between hierarchies. We, therefore, modify the classical simplification \cite{garland1997surface} to simplify mesh more uniform and record the correspondences between coarse mesh and dense mesh for newly defined mesh pooling and de-pooling operations.}

\noindent\textbf{Deforming Mesh Representations and Applications.} 
In order to better represent 3D meshes, a straightforward approach is to use vertex coordinates of a 3D shape. However, vertex coordinates are neither translation invariant nor rotation invariant, making it difficult to learn large-scale deformation. We instead use a recent 3D shape deformation representation~\cite{gao2017sparse}, which compared with another widely used representation~\cite{Gao:2016:EFD}, has the benefit of recording deformations at vertices, making graph convolutions and pooling operations easier to achieve.

Shape generation and interpolation are common applications on mesh data. Utilizing the VAE structure, MeshVAE~\cite{meshVAE} generates more deformable mesh shapes, and the method in \cite{ranjan2018generating} generates 3D faces with vivid expressions from a latent space. In fact, these VAE-based methods~\cite{meshVAE,ranjan2018generating,Litany_2018_CVPR} can also be used for the task of shape interpolation. Shape interpolation is a well researched topic. Existing mesh interpolation methods can be largely categorized into geometry-based methods (e.g.~\cite{huber2017smooth}) and data-driven methods (e.g.~\cite{CGF:CGF12991}). The latter exploits the information hidden in the shape dataset, and thus such approaches can produce more reasonable and reliable interpolation results. It has been shown that meshVAE~\cite{meshVAE} achieves better results than existing data-driven methods~\cite{CGF:CGF12991}. However, \cite{meshVAE} cannot deal with the shapes that contain too many vertices (e.g. the elephant model from \cite{sumner2004deformation}), which restricts the resolution of generated mesh shapes. Although the method in \cite{ranjan2018generating} performs well on face shapes, its reconstruction results on human body~\cite{vlasic2008articulated} are not satisfactory. 
Our work is also based on the VAE architecture, and therefore can naturally be used for shape generation and shape interpolation. Our framework improves over MeshVAE, and has better generalization ability, as we will demonstrate later.

\begin{figure*}
	\begin{center}
		\includegraphics[width=.9\linewidth]{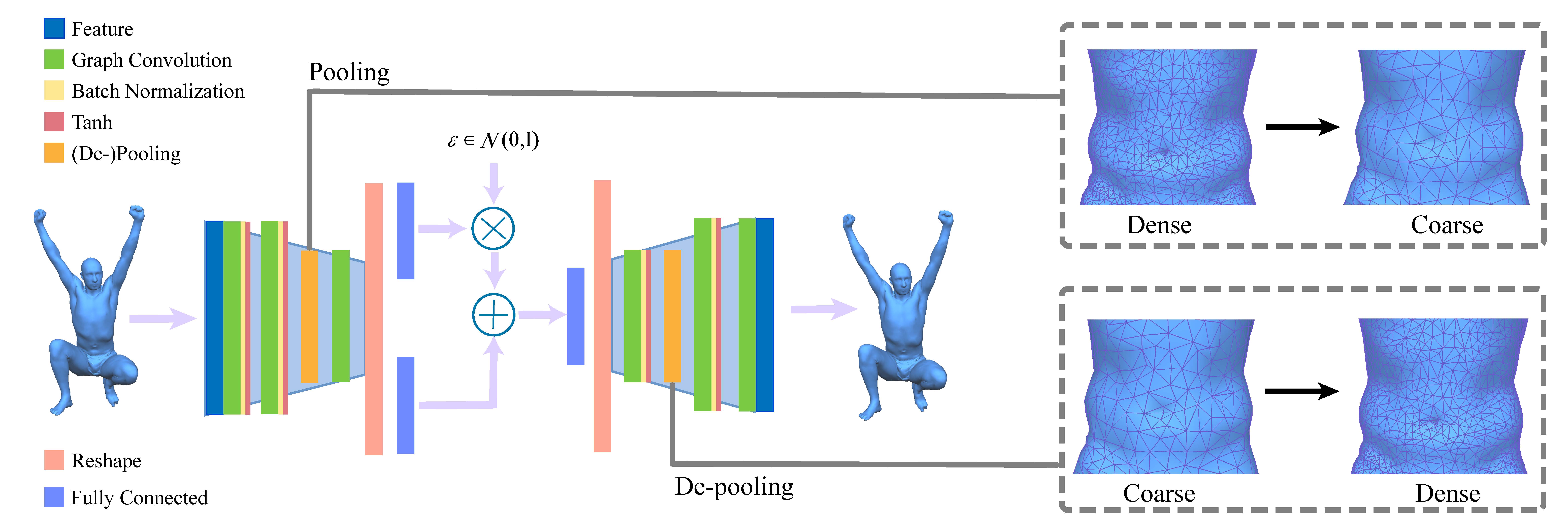}
	\end{center}\vspace{-2mm}
	\caption{Our network architecture. $\epsilon$ is a random variable with a Gaussian distribution with 0 mean and unit variance.}\vspace{-3mm}
	\label{fig:network_structure}
\end{figure*}

\section{Feature Representation}
To better represent shapes used in our network, as previously discussed, we utilize a recent deformation representation~\cite{gao2017sparse}. 
We assume that shapes in the dataset have the same mesh connectivity. This assumption is satisfied by many deformable object datasets \cite{anguelov2005scape,vlasic2008articulated,sumner2004deformation}, and the same connectivity can be achieved with consistent remeshing. Let $N$ be the number of shapes in the dataset, and the $m^{\rm th}$ shape is represented as $S_m$.
$\mathbf{p}_{m,i}\in\mathbb{R}^3$ is the $i^{\rm th}$ vertex on the $m^{\rm th}$ model. The deformation gradient $\mathbf{T}\in\mathbb{R}^{3\times3}$ is a local affine transform that describes local deformation around a vertex.
Using polar decomposition, $\mathbf{T}_{m,i}=\mathbf{R}_{m,i}\mathbf{S}_{m,i}$, that is, the deformation gradient $\mathbf{T}_{m,i}$ can be decomposed into a rotation matrix  $\mathbf{R}_{m,i}$ and a scale/shear matrix $\mathbf{S}_{m,i}$. 
By collecting non-trivial entries in the rotation and scale/shear components, the deformation around the $i^{\rm th}$ vertex of the $m^{\rm th}$ shape can be represented as a vector $q_{m,i}\in \mathbb{R}^9$. Following~\cite{meshVAE}, We further apply a linear scaling to map each element of $q_{m,i}$ to $[-0.95, 0.95]$ to allow $tanh$ to be used as an activation function. 

\section{Our Framework}
In this section we introduce the basic operations and network architecture used in our framework. We first describe our modified mesh simplification algorithm, which will be used to build mesh hierarchy for our pooling operations. Then the graph convolutional operation will be introduced. Finally we will summarize our network structure.

\subsection{Mesh Simplification}
\begin{figure}
	\begin{center}
		\includegraphics[width=1\linewidth]{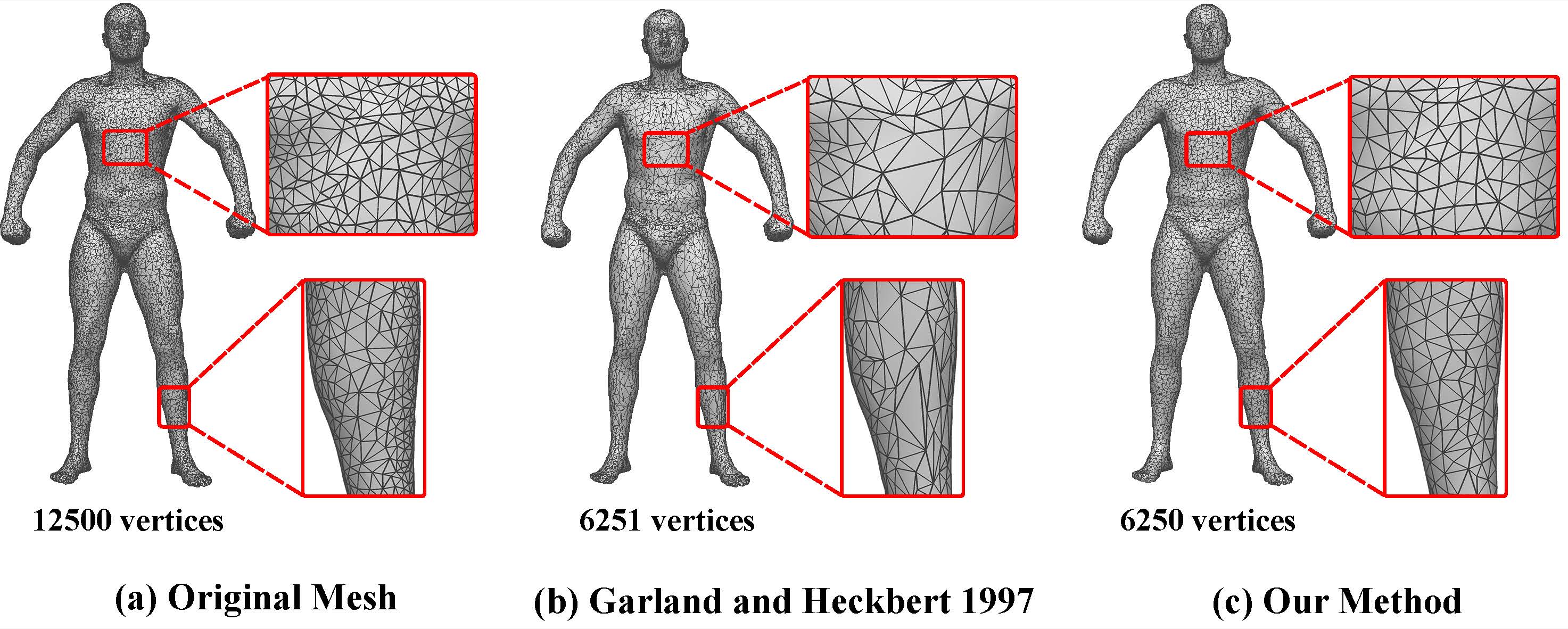}
	\end{center}\vspace{-2mm}
	\caption{Comparison of the mesh simplification algorithm~\cite{garland1997surface} and our modified version. (a) the original mesh with 12,500 vertices, (b) a result of~\cite{garland1997surface} with 6,251 vertices, and (c) our result with 6,250 vertices.}\label{fig:scape_sim}\vspace{-3mm}
\end{figure}

We use mesh simplification to help build reliable pooling operations.
For this purpose, mesh simplification not only creates a mesh hierarchy with different levels of details, but also ensures the correspondences between coarser and finer meshes. 
Our simplification process is based on the classical method~\cite{garland1997surface}, which performs repeated edge contraction in an order based on a metric measuring shape changes. However, the original approach cannot guarantee that the simplified mesh contains evenly distributed triangles. 
% \hongbo{Does this problem exist the follow-up work of \cite{garland1997surface}?} \yyja{We want the mesh pooling operation based on this kind of simplification can pooling the vertex feature evenly. "simplified mesh containing evenly distributed triangles" may not be a requirement in mesh simplification task. Other modified methods focus on how to preserve appearance and keep the quality of the mesh. \hongbo{I'm not sure if your argument is true or not, since making the vertices in a simplified mesh evenly distributed seems a good property for other applications. I would suggest you to conduct a thorough survey to better understand if we have made any contribution here.}}
To achieve more effective pooling, each vertex in the coarser mesh should correspond to a similarly sized region.

Our observation is that the edge length is an important indicator for this process. To avoid contracting long edges, we incorporate the edge length as one of the criteria to order pairs of points to be simplified. The original work defines the error at vertex $\mathbf{v} = [v_x, v_y, v_z, 1]^{\mathrm{T}}$ to be a quadratic form $\mathbf{v}^{\mathrm{T}}\mathbf{Q}\mathbf{v}$, {where $\mathbf{Q}$ is the sum of the fundamental error quadrics introduced in \cite{garland1997surface}}. 
% \hongbo{what is $\mathbf{Q}$?}
For a given edge contraction $(\mathbf{v}_1, \mathbf{v}_2)\rightarrow\bar{\mathbf{v}}$, they simply choose to use $\bar{\mathbf{Q}}=\mathbf{Q}_1+\mathbf{Q}_2$ to be the new matrix which approximates the error at $\bar{\mathbf{v}}$. So the error at $\bar{\mathbf{v}}$ will be $\bar{\mathbf{v}}^{\mathrm{T}}\bar{\mathbf{Q}}\bar{\mathbf{v}}$. 
%As we want the edge length to become one of the indicators, 
%we need to figure out the form in which it participate in the simplification. We finally choose to 
We propose to add the new edge length to the original simplification error metric. Specifically, given an edge $(\mathbf{v}_i, \mathbf{v}_j)$ to be contracted to a new vertex $\bar{\mathbf{v}}_k$,  the total error is defined as:
\begin{equation} \label{eq:simplification}
\begin{aligned}
E &= \bar{\mathbf{v}}_k^{\mathrm{T}}\bar{\mathbf{Q}}_k\bar{\mathbf{v}}_k \\
&+ \lambda \max\{L_{km}, L_{kn} |m \in \mathcal{N}_i, n \in \mathcal{N}_j, m \neq j, n \neq i \} ,
\end{aligned}
\end{equation}
where $L_{km}$ (resp. $L_{kn}$) is the new edge length between vertex $k$ and vertex $m$ (resp. vertex $n$). $\mathcal{N}_i$ (resp. $\mathcal{N}_j$) is the set of neighboring vertices of vertex $i$ (resp. vertex $j$), and $\lambda$ is a weight. %As written in the expression, we only impose constraints on the maximum length of the new edges, which ensures that our revision is more goal-oriented. 
Note that we only penalize the maximum edge length around newly created vertices $\bar{\mathbf{v}}_k$ to effectively avoid triangles with too long edges.
% This performs better than using the average of new edge lengths \hongbo{used in the \cite{garland1997surface}?}. 
%If we constrain the average of all new side lengths, the algorithm tends to be more unstable because more edges are involved; if a constraint is imposed on edge change ratio, then with the same ratio, the length of the original edge length will affect the actual value of edge change, which is what we do not want to see. 
In our experiments, we contract half of the vertices between adjacent levels of details to support effective pooling. 
% \hongbo{But in fact, we apply the pooling operation only once in the network? It might make better sense to use it for multiple times }.\yyja{We have tried to use the pooling operation for multiple times but got worser results.} \hongbo{well, this seems a quite serious limitation.}
A representative simplification example is shown in Fig.~\ref{fig:scape_sim}, which clearly shows the effect of our modified simplification algorithm.
The advantage of our modified simplification algorithm over the original one on pooling and thus shape reconstruction will be discussed in Section \ref{sec:framework_evaluation}.

%\hongbo{\cite{garland1997surface} is a very old technique. I'm sure there are more powerful/effective simplification methods. In addition, it is important to demonstrate the effect of the modified simplification method in terms of the pooling operation and target applications.}\yyja{We need a kind of edge contraction simplification algorithm, and \cite{garland1997surface} is a stable method and still being used in many applications, like OpenMesh. That's why we choose it. We show reconstruction error comparison in experiments.}

\begin{figure}
	\begin{center}
		\includegraphics[width=1\linewidth]{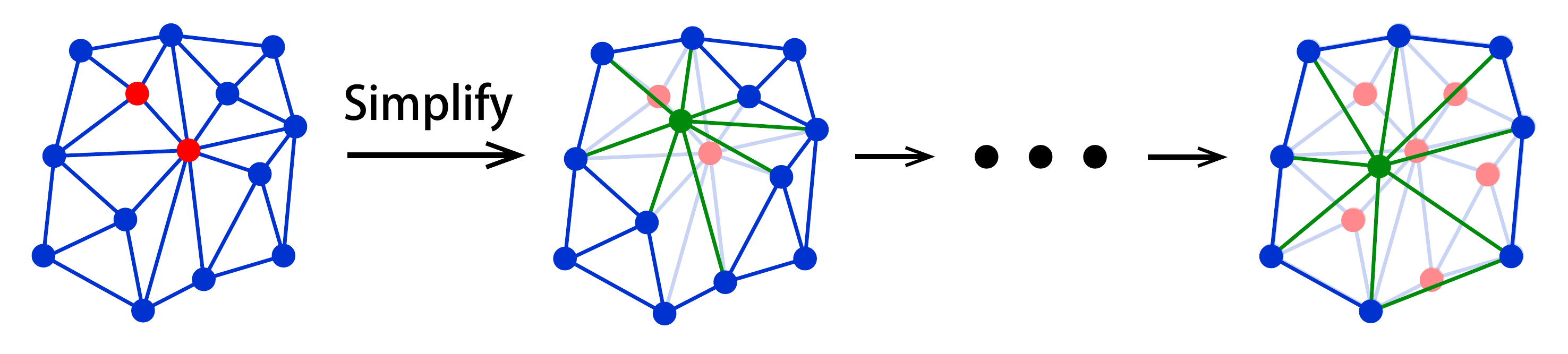}
	\end{center}
	\vspace{-2mm}
	\caption{We use a simplification algorithm to introduce our pooling operation on meshes. The red vertices are simplified to the green vertex by edge contraction and the features of the red vertices are averaged to give the feature of the green vertex.}\vspace{-3mm}
	\label{fig:pooling}
\end{figure}

\subsection{Pooling and De-pooling}\label{sec:pooling}
Mesh simplification is achieved by repeated edge contraction, i.e., contracting two adjacent vertices to a new vertex. 
%which means that two adjacent vertices have been contracted to one new vertex. After simplification, every vertex on the simplified mesh can build relationships with the vertex on the dense mesh. 
%We record those relationships and use them to do our pooling operation. The pooling operations we defined are brand new and have a similar feel to the pooling operations defined on the image. 
We exploit this process to define our pooling operation, in a way similar to image-based pooling. 
%The most important part is that we directly translate the simplification into a pooling operation. 
%If the points are randomly sampled from the mesh, this will destroy the topology, making further convolution operations impossible; if we look for the nearest point after simplification, the correspondence is not accurate or scientific.
%In our work, 
We use average pooling for our framework (and alternative pooling operations can be similarly defined). As illustrated in Fig. \ref{fig:pooling}, following an edge contraction step, we define the feature of a new vertex as the average feature of the contracted vertices. This ensures that the pooling operation effectively operates at relevant simplified regions. This process has some advantages: It preserves a correct topology to support multiple levels of convolutions/pooling, and makes the receptive field well defined. 
%regard the average of the vertex feature of contracted vertex as the feature of the vertex on simplified mesh. If the vertex on the dense mesh has been preserved after simplification, it will be pooled to itself and other vertex will not be pooled to it, that is, its feature will be kept.

% \begin{figure*}
% 	\begin{center}
% 		\includegraphics[width=.9\linewidth]{image/network_structure1.pdf}
% 	\end{center}\vspace{-2mm}
% 	\caption{Our network architecture. $\epsilon$ is a random variable with a Gaussian distribution with 0 mean and unit variance.}\vspace{-3mm}
% 	\label{fig:network_structure}
% \end{figure*}

Since our network has a decoder structure, we also need to properly define a de-pooling operation. We similarly take advantage of simplification relationships, and define de-pooling as the inverse operation: the features of the vertices on the simplified mesh are  equally assigned to the corresponding contracted vertices on the dense mesh.
% \hongbo{what are the new vertex's vertices?} 

\subsection{Graph Convolution}
To form a complete neural network architecture, we adopt 
% \hongbo{by using ``adapt'' it often means you have made some changes over the existing approach. If yes, please highlight the changes. Otherwise use ``adopt''?}\yyja{"adopt" is correct, I mix the meaning of these two words} 
the spectral graph convolutions introduced in \cite{defferrard2016convolutional}.
Let $\mathbf{x}$ be the input and $\mathbf{y}$ be the output of a convolution operation. $\mathbf{x}$ and $\mathbf{y}$ are matrices where each row corresponds to a vertex and each column corresponds to a feature dimension. Let $\mathbf{L}$ denote the normalized graph Laplacian. The spectral graph convolution used in our network is then defined as
\begin{equation}\label{eq:exp}
\mathbf{y} = g_{\theta}(\mathbf{L})\mathbf{x} = \sum_{h=0}^{H-1}{\theta_h \mathbf{T}_h(\tilde{\mathbf{L}})\mathbf{x}},
\end{equation}
where $\tilde{\mathbf{L}} = 2\mathbf{L}/\lambda_{max} - \mathbf{I}$, $\theta\in\mathbb{R}^H$ is polynomial coefficients, and $\mathbf{T}_h(\tilde{\mathbf{L}})\in\mathbb{R}^{V \times V}$ is the Chebyshev polynomial of order $h$ evaluated at $\tilde{\mathbf{L}}$.
%We empirically set the hyper-parameter $H = 3$ in our experiments. 

\subsection{Network Structure}
As illustrated in Fig. \ref{fig:network_structure}, {our overall network is built on our average pooling operation and  convolutional operation, with a variational auto-encoder structure.} The input to the encoder is the preprocessed features which are shaped as $X \in \mathbb{R}^{V \times 9}$, where $V$ is the number of vertices and 9 is the dimension of the deformation representation.

Unlike the original mesh VAE~\cite{meshVAE}, which uses fully connected layers, our approach benefits from graph convolutions and especially, the newly defined pooling operations to massively reduce the number of parameters.
% \hongbo{This is misleading, since graph convolutions have already been used for mesh VAE. You cannot oversell the contribution.} 
% We also introduce \yyj{newly defined} pooling operations to improve the generalizability of the network.
The introduction of the pooling operations also improves the generalizability of the network.

Our network takes the preprocessed vertex features defined in~\cite{gao2017sparse} as input, which then go through two graph convolutional layers, followed by one pooling layer and another graph convolutional layer. 
% 	The last convolutional layer directly uses linear output without any non-linear activation function to avoid overfitting. 
In order to avoid overfitting, the last convolutional layer does not use any nonlinear activation function.
The output of the last convolutional layer is mapped to a mean vector and a deviation vector by two different fully-connected layers. The mean vector does not have an activation function, and the deviation vector uses \emph{sigmoid} as the activation function. 
%\yyja{$\sigma_{max}$ is used for embedding, which is not a novel application, so we delete it.}
%which is then multiplied by an upper bound value $\sigma_{max}$ \hongbo{($\sigma_{max} = xxx$ in our experiments)} of the expected deviation. We use $\sigma_{max}=1$ in our experiments.

To reconstruct the shape representation from the latent vector, the decoder is used, which basically mirrors the encoder steps. For the decoder convolutional layers, we use the transposed weights of the corresponding layers in the encoder, with all layers using the $tanh$ output activation function. Corresponding to the pooling operation, the de-pooling operation as described in Section \ref{sec:pooling} maps features in a coarser mesh to a finer mesh. The output of the whole network is $\hat{X}\in\mathbb{R}^{V \times 9}$, which has the identical dimension as the input, and can be rescaled back to the deformation representation and used for reconstructing the deformed shape.

In order to train the model to fit the large dimension of mesh features, we use the mean squared error (MSE) as the reconstruction loss. Combined with the KL-divergence \cite{kullback1951information}, the total loss function for the model is defined as
% we replace the reconstruction loss from a probabilistic formulation \yyj{\cite{kingma2013auto}} 
% \hongbo{used in existing work? reference is needed} 
% with a simpler mean square error (MSE), \yyj{as \cite{meshVAE} did}. \hongbo{you may also want to show the advantage with the changed loss} 
% The total loss function for the model is defined as
\begin{equation}
L = \frac{1}{2M} \sum_{i=1}^M {\|X^i-\hat{X}^i\|^2_F} + \alpha D_{KL}(q(z|X)\|p(z)),
\end{equation}
where $X^i$ and $\hat{X}^i$ represent the preprocessed features of the $i^{\rm th}$ model and the output of the network. $\|\cdot\|_F$ is the Frobenius norm of matrix, $M$ is the number of shapes in the dataset, $\alpha$ is a parameter to adjust the priority between the reconstruction loss and KL-divergence. $z$ is the latent vector, $p(z)$ is the prior probability, $q(z|X)$ is the posterior probability, and $D_{KL}$ is the KL-divergence. 
%\hongbo{what is $||$ in the second term?}\yyja{It's a symbol in KL-divergence}
%The pipeline for our framework is summarized in Fig.~\ref{fig:network_structure}.

\subsection{Conditional VAE}
When the VAE is used for shape generation, it is often preferred to allow the selection of shape types to be generated, especially for datasets containing shapes from different categories (such as men and women, thin and fat, see~\cite{pons2015dyna} for more examples). To achieve this, we refer to~\cite{sohn2015learning} and add labels to the input and the latent vectors to extend our framework. In this case, our loss function is changed to 
\begin{equation}
\begin{aligned}
&L_{c} = \frac{1}{2M} \sum_{i=1}^M {\|X_{c}^i-\hat{X}^i\|^2_F} + \alpha D_{KL}(q(z|X,c)\|p(z|c)),
\end{aligned}
\end{equation}
where {$\hat{X}$} is the output of the conditional VAE, and $p(z|c)$ and $q(z|X,c)$ are conditional prior and posterior probabilities, respectively. %We present our experimental results in the following section.

\subsection{Implementation Details}
In our experiments, we contract half of the vertices with $\lambda=0.001$ in Eq.~\ref{eq:simplification} and set the hyper-parameter $H=3$ in graph convolutions, $\alpha=0.3$ in the total loss function. The latent space dimension is 128 for all our experiments. We also use $L_2$ regularization on the network weights to avoid over-fitting. We use Adam optimizer~\cite{kingma2014adam} with the learning rate set to $0.001$.

% \begin{table}
% 	\centering
% 	\begin{tabular}{|c|c|c|c|}
% 		\hline
% 		Dataset & Only Spatial Conv. & No Pooling & Our Method \\
% 		\hline
% 		SCAPE & 0.1086 & 0.0825 & \textbf{0.0763} \\
% 		\hline
% 		Swing & 0.0359 & 0.0282 & \textbf{0.0268} \\
% 		\hline
% 		Fat & 0.0362 & 0.0267 & \textbf{0.0249} \\
% 		\hline
% 		Hand & 0.0300 & 0.0284 & \textbf{0.0260} \\
% 		\hline
% 	\end{tabular}
% %	\vspace{1mm}
% 	\caption{Comparison of RMS errors for reconstructing unseen data using our network with pooling, without pooling and with different convolutional operators on several datasets. We use the network architecture as illustrated in Fig.~\ref{fig:network_structure}.}\vspace{-2mm}
% 	\label{table:ablation}
% \end{table}

\begin{table*}
	%	\vspace{-15mm}
	\centering
	\begin{tabular}{|c|c|c|c|c|c|c|c|}
		\hline
		\multirow{2}{*}{Dataset} & Only & Only & \multirow{2}{*}{\cite{garland1997surface}} & Uniform & Graph & Mesh & Our \\
		& Spatial Conv. & Spectral Conv. & & Simp. & Pooling & Sampling & Method \\
		\hline
		SCAPE & 0.1086 & 0.0825 & 0.0898 & 0.0813 & 0.0824 & 0.0831 & \textbf{0.0763} \\  
		\hline 
		Swing & 0.0359 & 0.0282 & 0.0284 & 0.0281 & 0.0292 & 0.0298 & \textbf{0.0268} \\  
		\hline
		Fat & 0.0362 & 0.0267 & 0.0285 & 0.0305 & 0.0253 & 0.0289 & \textbf{0.0249} \\  
		\hline
		Hand & 0.0300 & 0.0284 & 0.0271 & 0.0280 & 0.0306 & 0.0278 & \textbf{0.0260} \\  
		\hline
	\end{tabular}
	\vspace{1mm}
	\caption{\yyj{Comparison of RMS reconstruction errors for unseen data using our network with pooling, without pooling, with different convolutional operators, with original simplification \cite{garland1997surface}-based pooling, with uniform simplification-based pooling (remeshing \cite{remesh}), with graph pooling \cite{shen2018mining} and with mesh sampling \cite{ranjan2018generating}. Note that 'Only Spectral Conv.' also means 'No Pooling'.}}\vspace{-2mm}
	\label{table:different_sim}
\end{table*}

\begin{table}
	%\vspace{-7mm}
	\centering
	\begin{tabular}{|c|c|c|c|}
		\hline
		Dataset & `CPCPC' & `CCPCCPC' & Our Network \\
		\hline
		SCAPE & 0.0942 & 0.0863 & \textbf{0.0763} \\
		\hline
	\end{tabular}
	\vspace{1mm}
	\caption{Comparison of different network structures. `C' and `P' refer to a graph convolutional layer and a pooling layer respectively.
	}\vspace{-3mm}
	\label{table:structure_compare}
\end{table}

\begin{table}\footnotesize
	%\vspace{-5mm}
	\centering
	\begin{tabular}{|c|c|c|c|c|c|}
		\hline
		\multirow{2}{*}{Dataset} & \multirow{2}{*}{\#. Vertices} & Tan  & Gao & Ranjan & \multirow{2}{*}{Ours}  \\
		& & 2018 & 2018 & 2018 & \\
		\hline
		SCAPE & 12500 & - & 0.1086 & 0.1095 & \textbf{0.0763} \\
		\hline
		Swing & 9971 & - & 0.0359 & 0.0557 & \textbf{0.0268} \\
		\hline
		Fat & 6890 & 0.0308 & 0.0362 & 0.0324 & \textbf{0.0249} \\
		\hline
		Hand & 3573 & 0.0362 & 0.0300 & 0.0632 & \textbf{0.0260} \\
		\hline
		Face & 11849 & - & 1.0619 & 1.1479 & \textbf{0.7257} \\
		\hline
		Horse & 8431 & - & 0.0128 & 0.0510 & \textbf{0.0119} \\
		\hline
		Camel & 11063 & - & 0.0134 & 0.0265 & \textbf{0.0115} \\
		\hline
	\end{tabular}
	\vspace{1mm}
	\caption{Comparison of RMS reconstruction errors for unseen data using different auto-encoder frameworks proposed by Tan et al. ~\cite{meshVAE}, Gao et al. \cite{gao2018automatic}, and Ranjan et al. \cite{ranjan2018generating}. `-' means the corresponding method runs out of memory (largely due to the use of fully connected networks).}
	\label{table:ERMS}
\end{table}

\begin{table}
	%\vspace{-5mm}
	\centering
	\begin{tabular}{|c|c|c|c|}
		\hline
		Dataset & \#. Vertices & Tan et al. 2018 &  Ours \\
		\hline
		Fat & 6890 & 129,745,920 & \textbf{7,941,042} \\
		\hline
		Hand & 3573 & 68,610,048 & \textbf{4,118,706} \\
		\hline
	\end{tabular}
	\vspace{1mm}
	\caption{Comparison of parameters number with \cite{meshVAE}. It can be seen that our network require far fewer parameters than \cite{meshVAE}.}\vspace{-2mm}
	\label{table:parameters}
\end{table}

\section{Experiments} \label{sec:exp}
\subsection{Framework Evaluation}\label{sec:framework_evaluation} 
To compare different network structures and settings, we use several shape deformation datasets, including
SCAPE dataset~\cite{anguelov2005scape}, Swing dataset~\cite{vlasic2008articulated}, Face dataset~\cite{neumann2013sparse}, Horse and Camel dataset~\cite{sumner2004deformation}, Fat (ID:50002) from the MPI DYNA dataset~\cite{pons2015dyna}, and Hand dataset.
For each dataset, it is randomly split into halves for training and testing. We test the capability of the network to generate unseen shapes, and report the average RMS (root mean squared) errors. 

\noindent\textbf{Effect of Pooling}. 
In Table~\ref{table:different_sim} (Column 3 and 8) we compare the RMS errors of reconstructing unseen shapes with and without pooling. The RMS error is lower by an average of $8.36\%$ with pooling. The results show the benefit of our pooling and de-pooling operations. 

\noindent\textbf{Comparison with Spatial Convolutions.} We compare spectral graph convolutions with alternative spatial convolutions, both with a similar network architecture as shown in Fig.~\ref{fig:network_structure}. The comparison results are shown in Table~\ref{table:different_sim} (Column 2 and 3). One can easily find that spectral graph convolutions give better results. 

\yyj{\noindent\textbf{Comparisons with Other Pooling or Sampling Methods.} To demonstrate the benefit of our simplification-based pooling operation, we compare our pooling with with the original algorithm~\cite{garland1997surface} for pooling, the existing graph pooling method~\cite{shen2018mining}, and the mesh sampling operation introduced in \cite{ranjan2018generating}. What's more, we show comparisons to a representative uniform simplification method based on remeshing \cite{remesh}. This method is able to distribute vertices uniformly but loses geometry details. In contrast, our method aims for a uniform, and also shape-preserving simplification, which leads to better results. The results are shown in Table~\ref{table:different_sim}. The RMS error of our pooling for unseen data is lower by an average of $9.17\%$ compared to \cite{garland1997surface}-based pooling, $8.06\%$ compared to graph pooling~\cite{shen2018mining}, and $9.64\%$ compared to mesh sampling \cite{ranjan2018generating},  which shows our modified simplification algorithm is more effective in term of pooling and our pooling is superior on multiple datasets, leading to better generalization ability.}

%\noindent\textbf{Simplification Algorithms.} We compare our modified mesh simplification algorithm with the original algorithm~\cite{garland1997surface} for pooling, while keeping other settings unchanged.
%As shown in Table~\ref{table:different_sim}, with our modified algorithm, the RMS error is lower by an average of $9.17\%$, which shows our modified algorithm is more effective in term of pooling, leading to better generalization ability. 
%
%\noindent\textbf{Comparison with Graph Pooling.} 
%To demonstrate the benefit of our simplification-based pooling operation, we compare it with the existing graph pooing method~\cite{shen2018mining}. As shown in Table~\ref{table:different_sim}, our pooling is superior on multiple datasets: the RMS error is lower by an average of $8.06\%$.
%
%\noindent{\textbf{Comparison with Mesh Sampling.} 
%Further, we replace our pooling operation by the sampling operation introduced in \cite{ranjan2018generating} to show the benefit of our pooling operation. 
%Compared to the sampling operation, we adopt the modified simplification algorithm, and different procedures when changing the connectivity of the mesh.
%As shown in Table~\ref{table:different_sim}, on average, our pooling achieves $9.64\%$ lower RMS reconstruction errors on multiple datasets.}

\begin{figure}
	%\vspace{-10mm}
	\begin{center}
		\includegraphics[width=.8\linewidth]{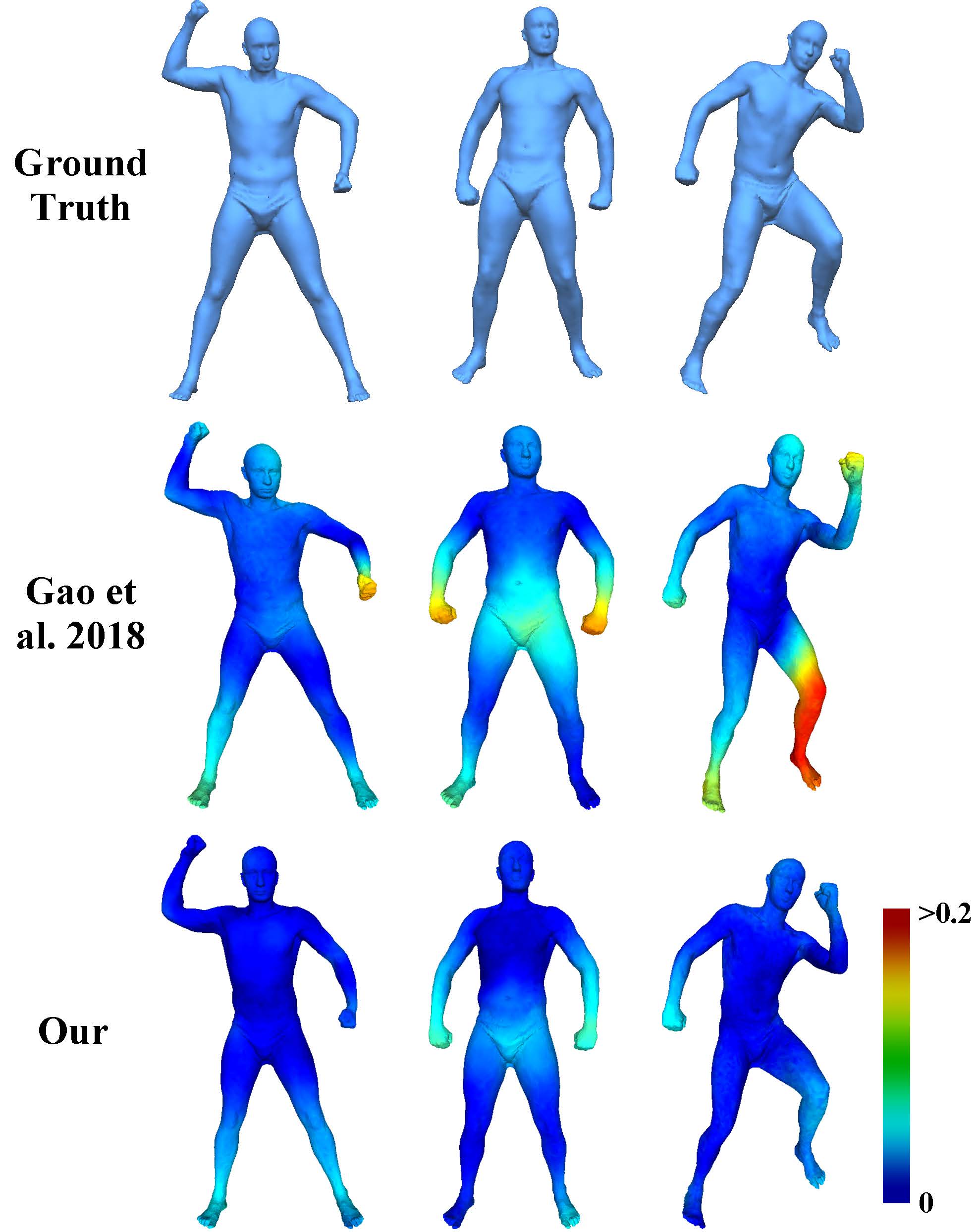}
	\end{center}
	\vspace{-2mm}
	\caption{Qualitative comparison of reconstruction results for unseen data with ~\cite{gao2018automatic}. Reconstruction errors are color-coded. It can be seen that our method leads to more accurate reconstructions.}\vspace{-2mm}
	\label{fig:recon_gao_comp}
\end{figure}

\begin{figure}
	%\vspace{-10mm}
	\begin{center}
		\includegraphics[width=.8\linewidth]{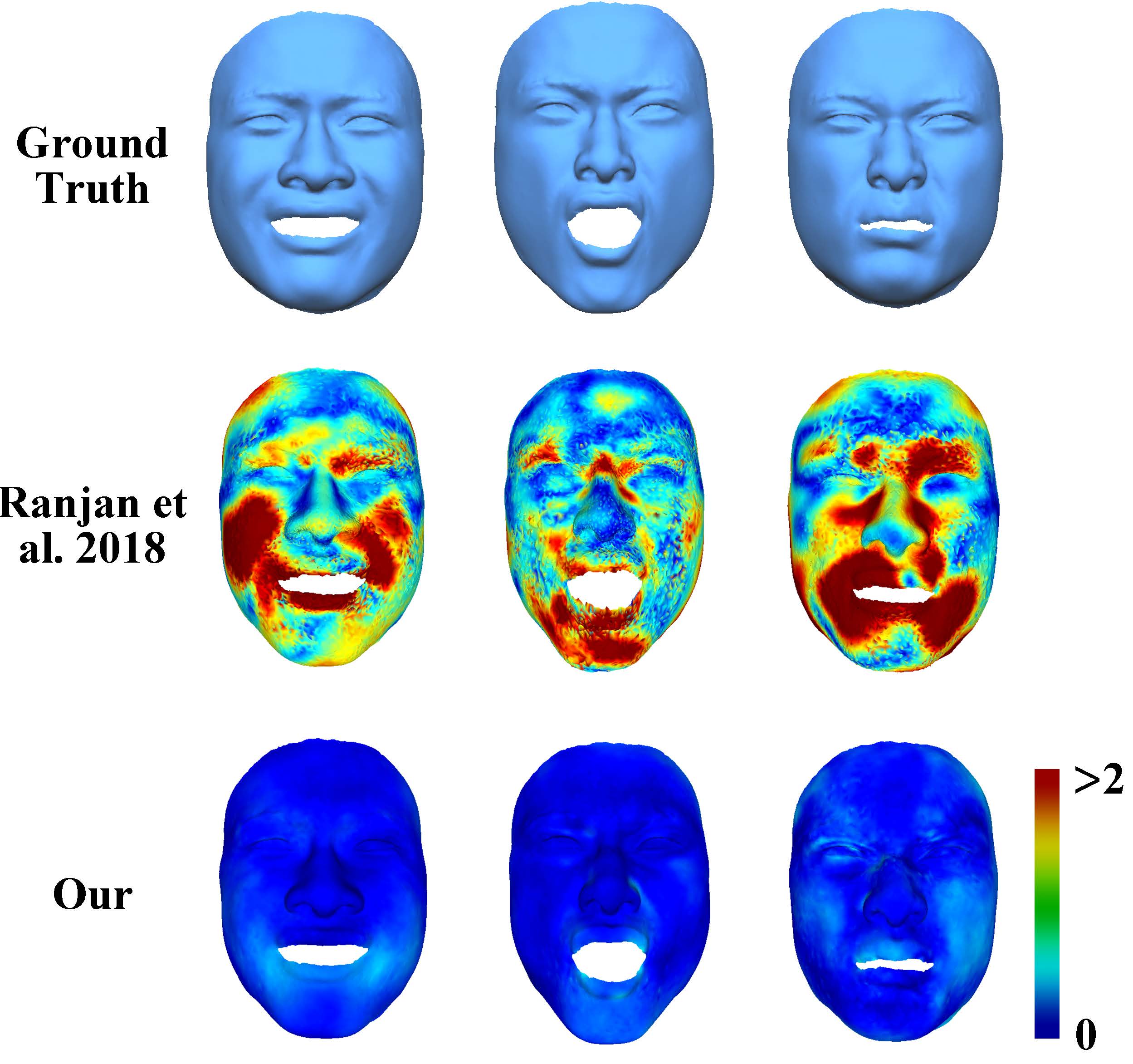}
	\end{center}
	\vspace{-2mm}
	\caption{Qualitative comparison of reconstruction results for unseen data with ~\cite{ranjan2018generating}. Our method leads to significantly lower reconstruction errors.}\vspace{-2mm}
	\label{fig:recon_coma_comp}
\end{figure}

\begin{figure}
	%\vspace{-10mm}
	\begin{center}
		\includegraphics[width=.8\linewidth]{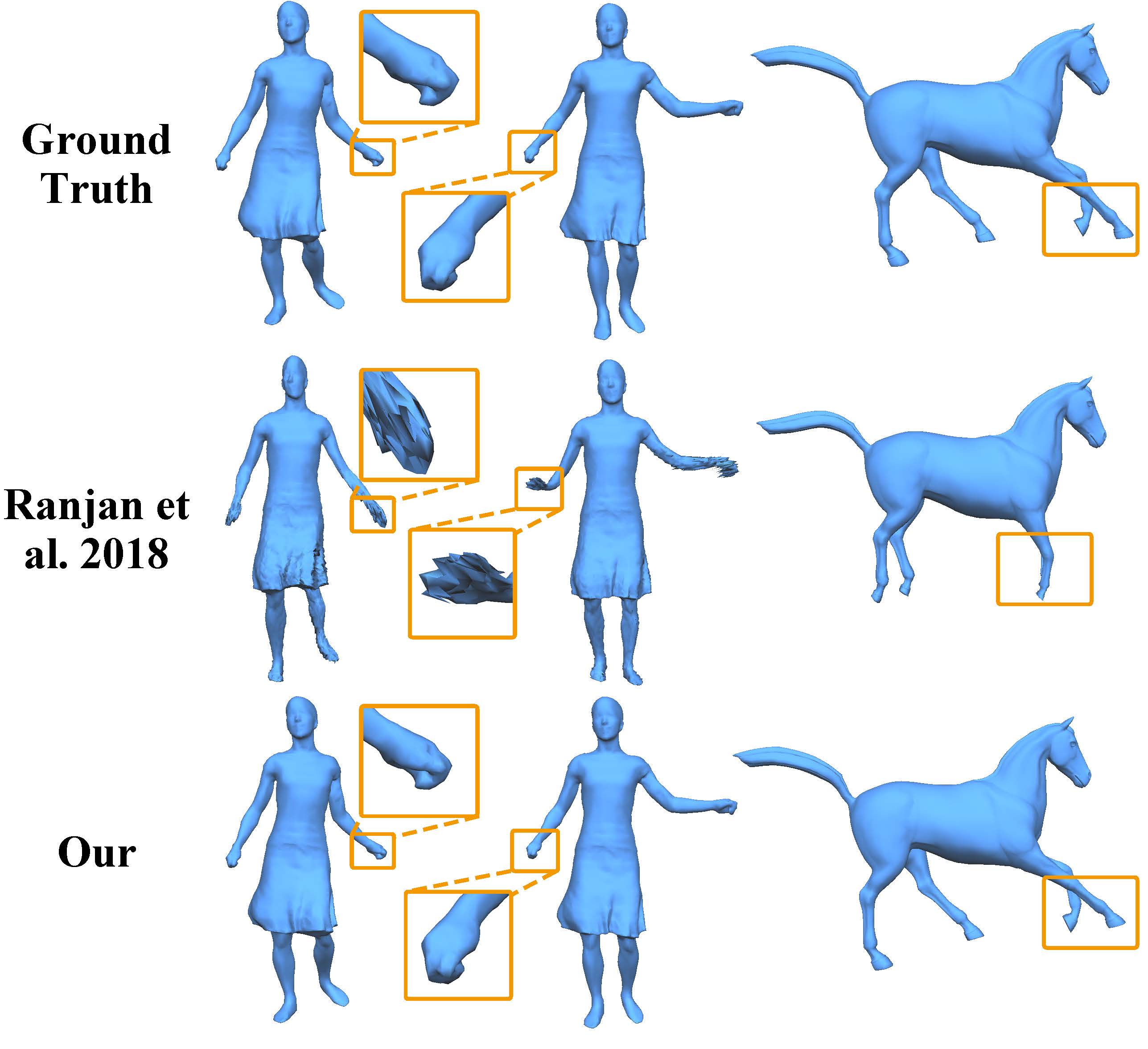}
	\end{center}
	\vspace{-2mm}
	\caption{{Qualitative comparison of reconstruction results with ~\cite{ranjan2018generating}. The method of \cite{ranjan2018generating} suffers from easily noticeable artifacts.}}\vspace{-2mm}
	\label{fig:recon_coma_comp1}
\end{figure}

\begin{figure*}
	\centering
	\includegraphics[width=.8\linewidth]{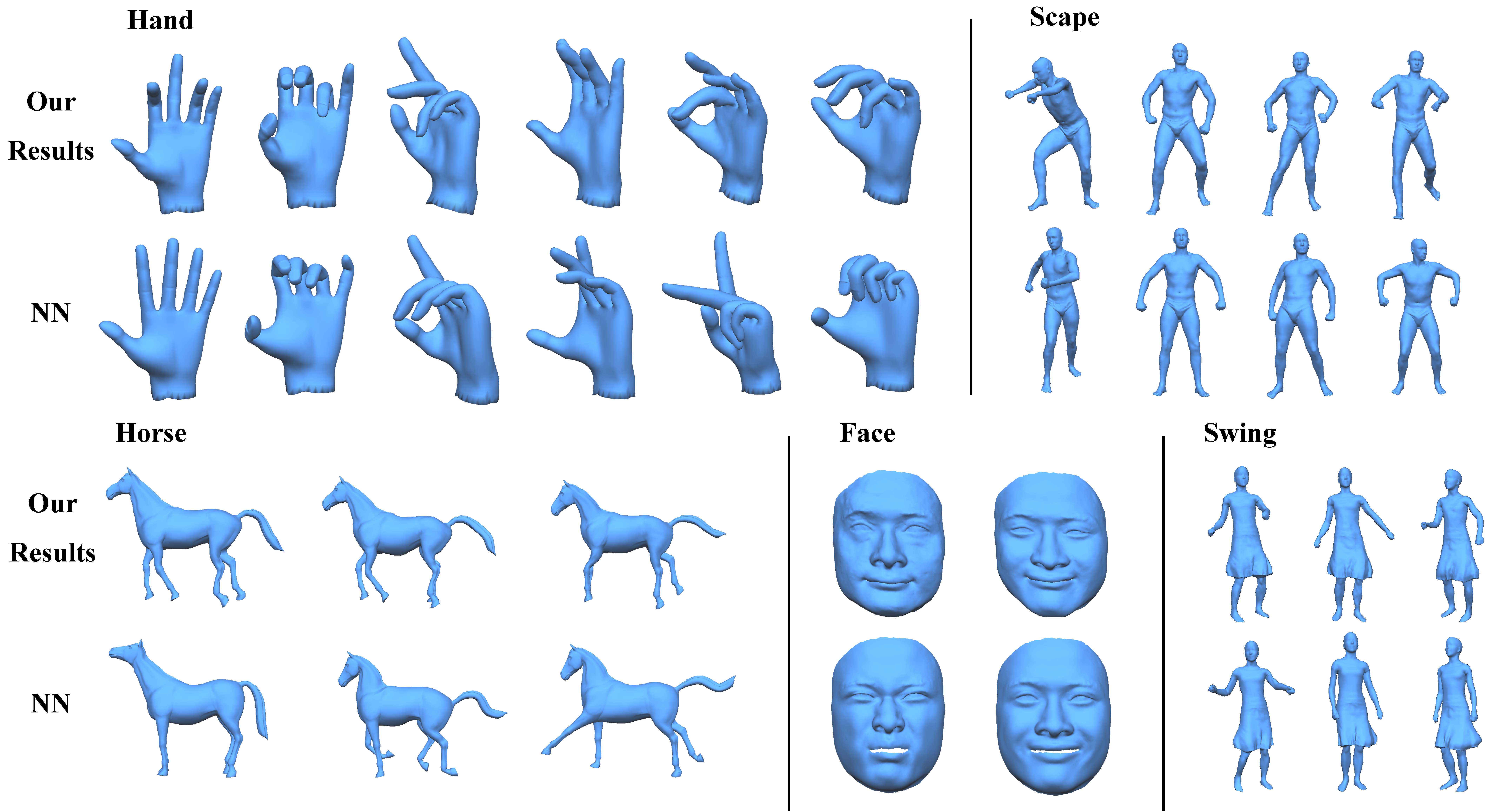}
	%\vspace{-2mm}
	\caption{Randomly generated new shapes using our framework, along with their nearest neighbors (NN) in the original datasets.}
	%\vspace{-3mm}
	\label{fig:randomly_generation}
\end{figure*}

\begin{figure}
	\begin{center}
		\includegraphics[width=.8\linewidth]{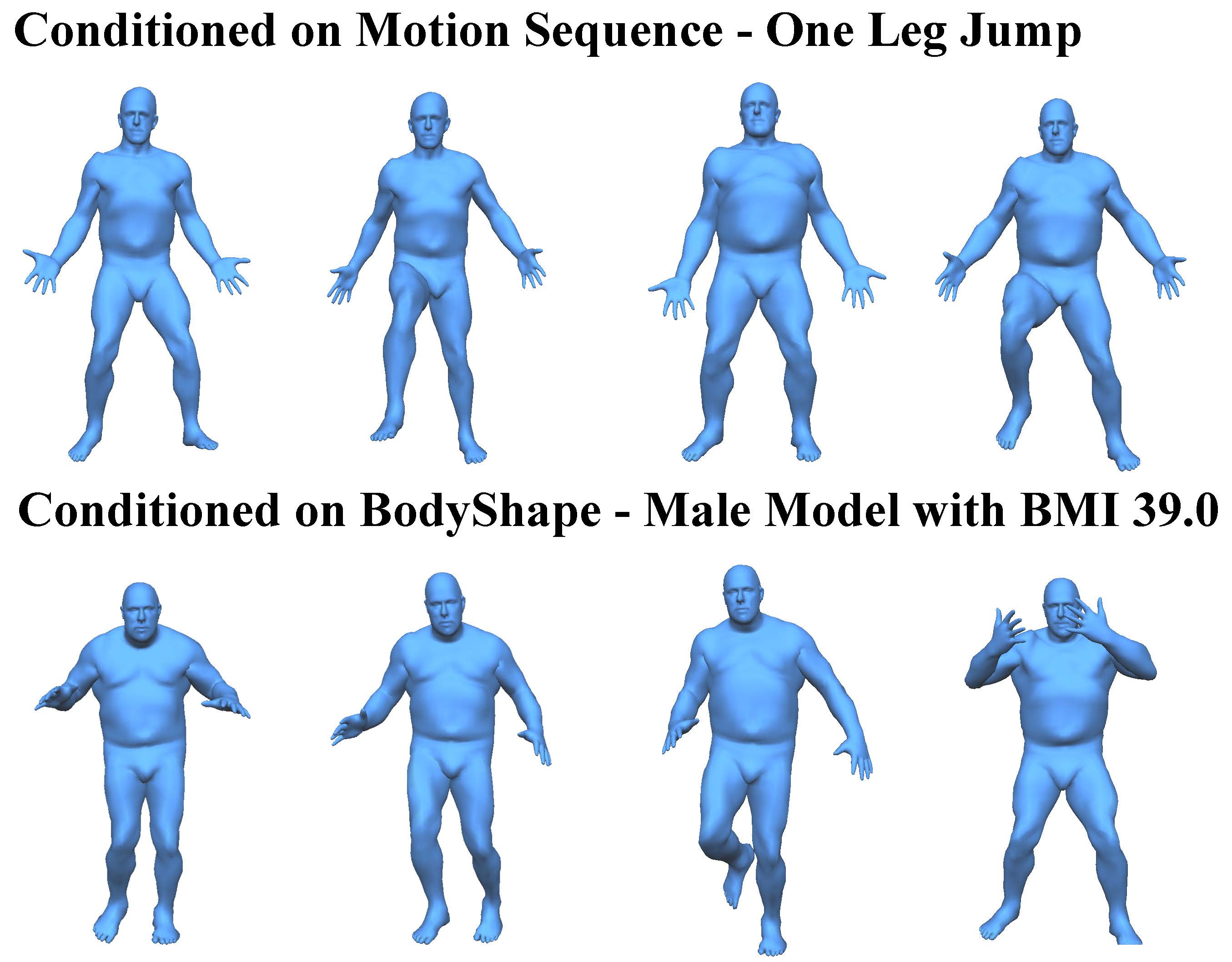}
	\end{center}
	%\vspace{-2mm}
	\caption{Conditional random generation of new shapes using our framework.}%\vspace{-3mm}
	\label{fig:conditional_generation}
\end{figure}

\noindent\textbf{Alternative Network Structures.} 
In addition to the architecture shown in Fig.~\ref{fig:network_structure}, we also consider alternative architectures, including adding more convolutional and pooling layers (see Table~\ref{table:structure_compare}). It can be seen that our chosen architecture has better performance than alternative architectures. This shows that our architecture is flexible enough to learn from the data but not overly complicated which might lead to overfitting.

\noindent\textbf{Comparison with State-of-the-Art.} In Table~\ref{table:ERMS},
% \hongbo{In this table you show the results on all the datasets. That's good. But the reviewer might ask why selected datasets are used for other comparisons/tables.}
we compare our method with the state-of-the-art mesh-based auto-encoder architectures~\cite{gao2018automatic,ranjan2018generating,meshVAE} in terms of RMS errors of reconstructing unseen shapes. 
{Thanks to spectral graph convolutions and our pooling}, our method consistently reduces the reconstruction errors of unseen data, showing superior generalizability. {For example, compared with \cite{gao2018automatic}, which uses the same per-vertex features as ours, our network achieves $29\%$ and $32\%$ lower average RMS reconstruction errors on the SCAPE and Face datasets. What's more, we show the qualitative reconstruction comparison with \cite{gao2018automatic} and \cite{ranjan2018generating} in Fig.~\ref{fig:recon_gao_comp}, \ref{fig:recon_coma_comp} and \ref{fig:recon_coma_comp1}. These figures show that our method leads to more accurate reconstruction results than \cite{gao2018automatic,ranjan2018generating}. In Table~\ref{table:parameters}, we present two comparisons to illustrate that our network requires far fewer parameters than the original MeshVAE.}
% Table~\ref{table:STED} further shows that our network outperforms the state-of-the-art variational auto-encoder structure~\cite{gao2018automatic} in terms of the STED perceptual metric.

\begin{figure}
	%\vspace{-10mm}
	\begin{center}
		\includegraphics[width=1\linewidth]{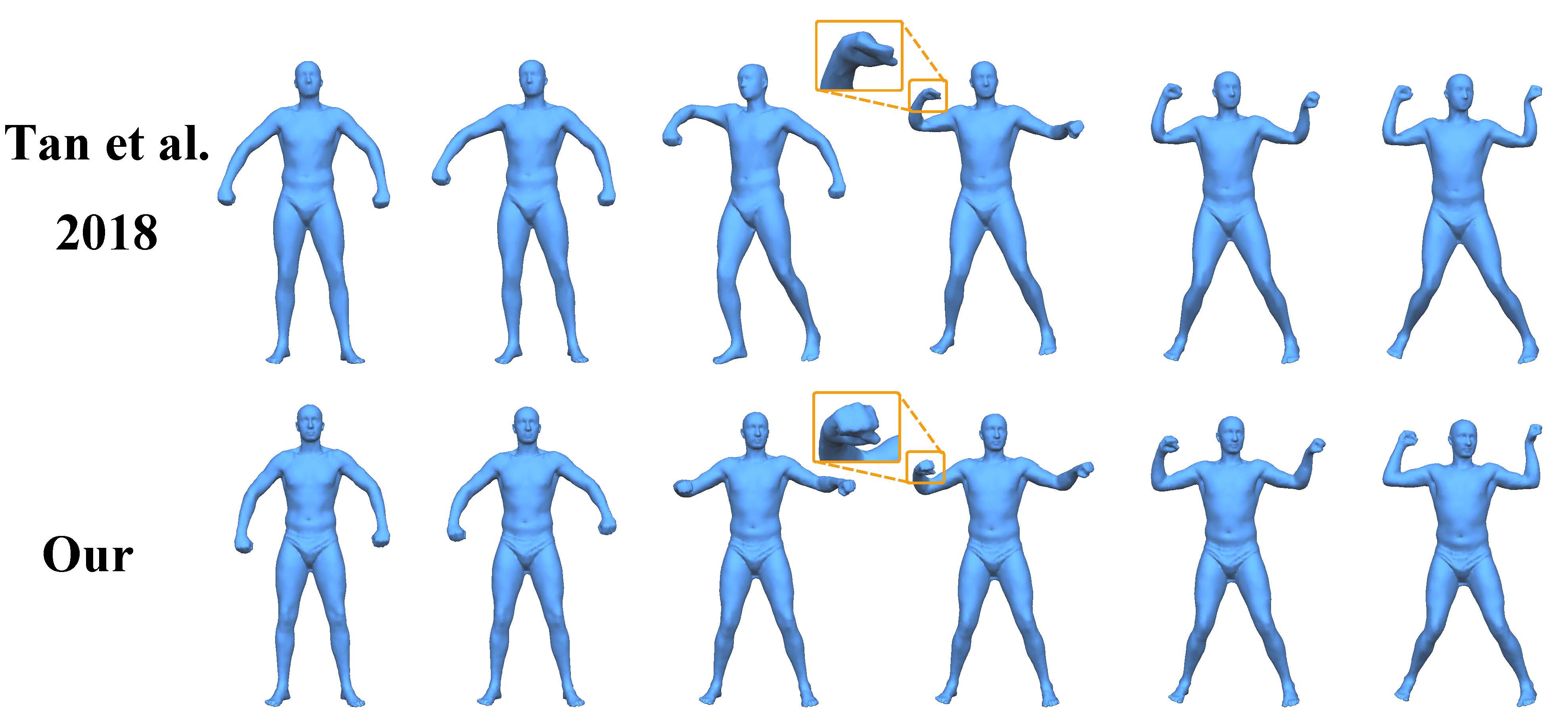}
	\end{center}
	\vspace{-2mm}
	\caption{Comparison of mesh interpolation results with~\cite{meshVAE}. The models in the leftmost and rightmost columns are the input models to be interpolated. 1st row: the results of~\cite{meshVAE}, and 2nd row: our results.}\vspace{-2mm}
	\label{fig:meshvae_comp}
	\vspace{-2mm}
\end{figure}

\subsection{Generation of Novel Models}
Once our network is trained, we can use the latent space and decoder to generate new shapes. We use the standard normal distribution $z \sim N(0,I)$ as the input to the trained decoder. %, and test the generation capability of mesh VAE++. 
%We train our network on the SCAPE dataset ~\cite{anguelov2005scape}, Swing dataset from ~\cite{vlasic2008articulated}, face dataset, Horse dataset ~\cite{sumner2004deformation} and hand dataset. 
It can be seen from Fig.~\ref{fig:randomly_generation} that our network is capable of generating reasonable new shapes. To prove that the generated shapes do not exist in the model dataset, we find the nearest shapes based on the average per-vertex Euclidean distance in the original datasets for visual comparison. 
%The way we look for the nearest neighbors is to look for the smallest Euclidean distance with the models in the dataset. 
%It can be seen from  that there are differences between our newly generated model and the nearest model from the dataset. 
It can be seen that the generated shapes are indeed new and different from any existing shape in the datasets. To show our conditional random generation ability, we train the network on the DYNA dataset from~\cite{pons2015dyna}. We use BMI+gender and motion as the condition to train the network. As shown in  Fig.~\ref{fig:conditional_generation}, our method is able to randomly generate models that are conditioned on the body shape `50007' -- a male model with BMI 39.0 and conditioned on the action with the label `One Leg Jump' including lifting the leg.

\begin{figure}
	%\vspace{-5mm}
	\begin{center}
		\includegraphics[width=1\linewidth]{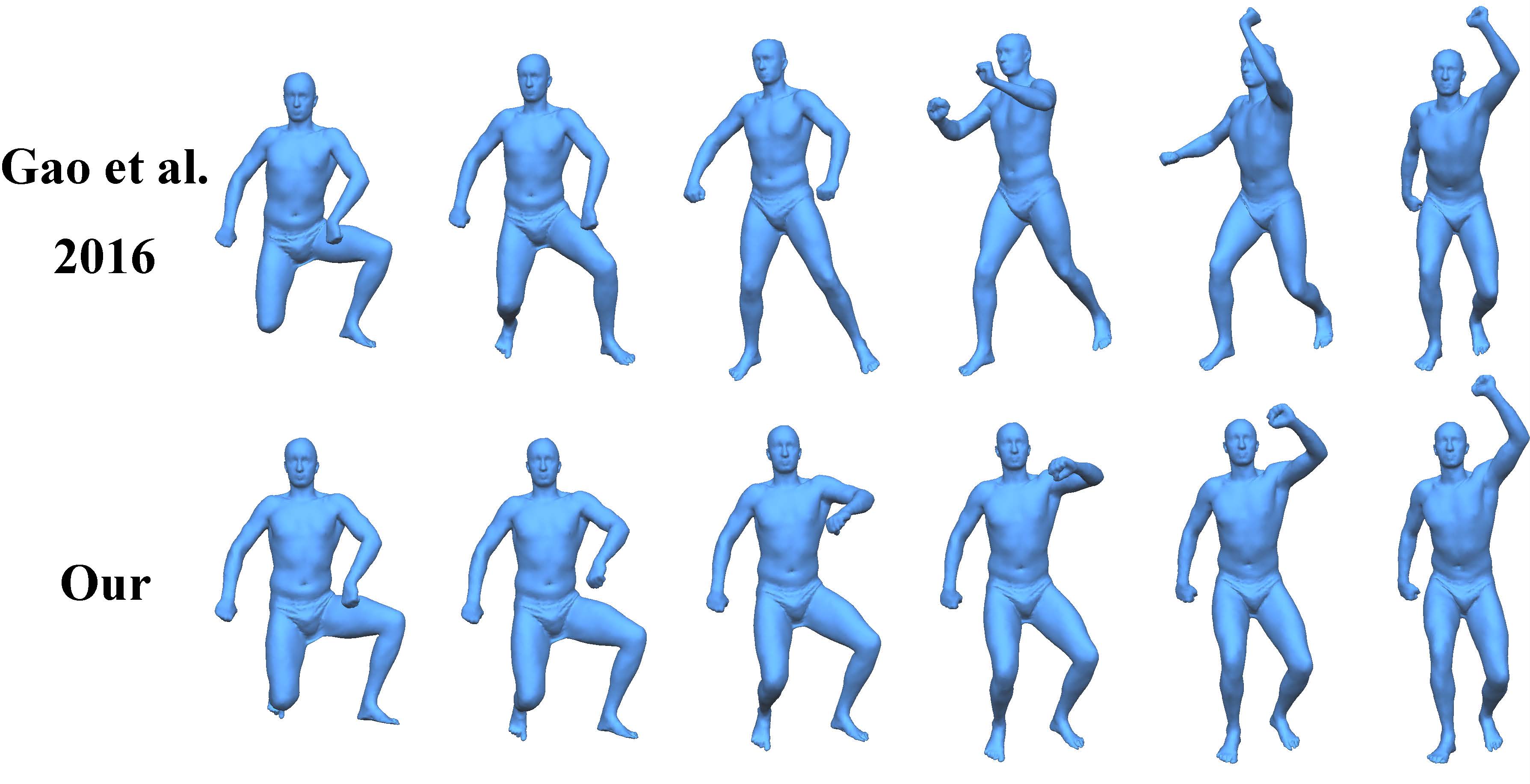}
	\end{center}
	\vspace{-2mm}
	\caption{Comparison of mesh interpolation results with~\cite{CGF:CGF12991} (1st row). The models in the leftmost and rightmost columns are the input models to be interpolated.}\vspace{-3mm}
	\label{fig:DD_comp}
\end{figure}

\begin{figure}
	%\vspace{-5mm}
	\begin{center}
		\includegraphics[width=1\linewidth]{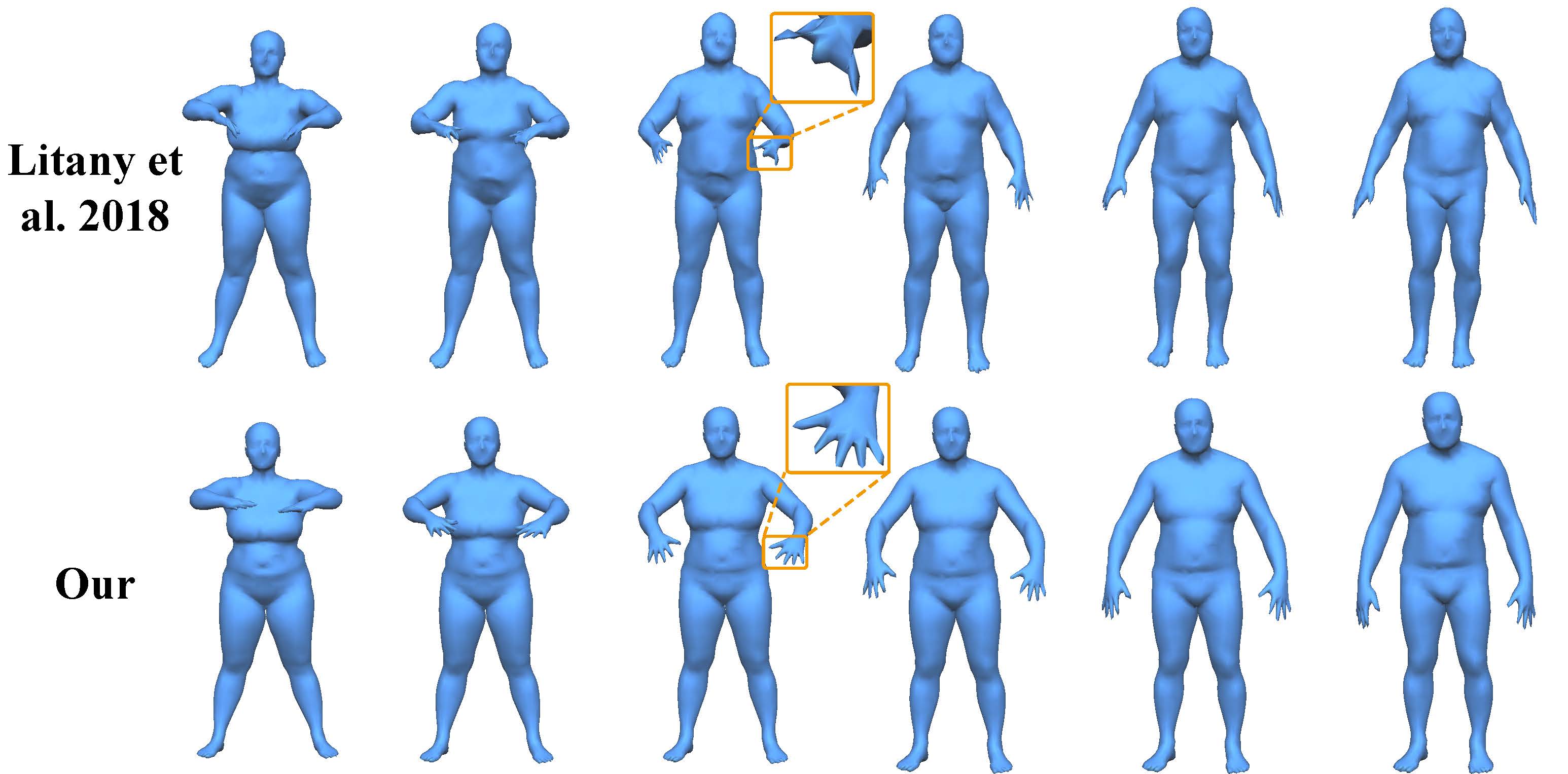}
	\end{center}
	\vspace{-2mm}
	\caption{Comparison of mesh interpolation results with~\cite{Litany_2018_CVPR}. First row is the result of~\cite{Litany_2018_CVPR}, and second row is our result. }\vspace{-5mm}
	\label{fig:litany_comp}
\end{figure}

\begin{figure}
	%\vspace{-5mm}
	\begin{center}
		\includegraphics[width=1\linewidth]{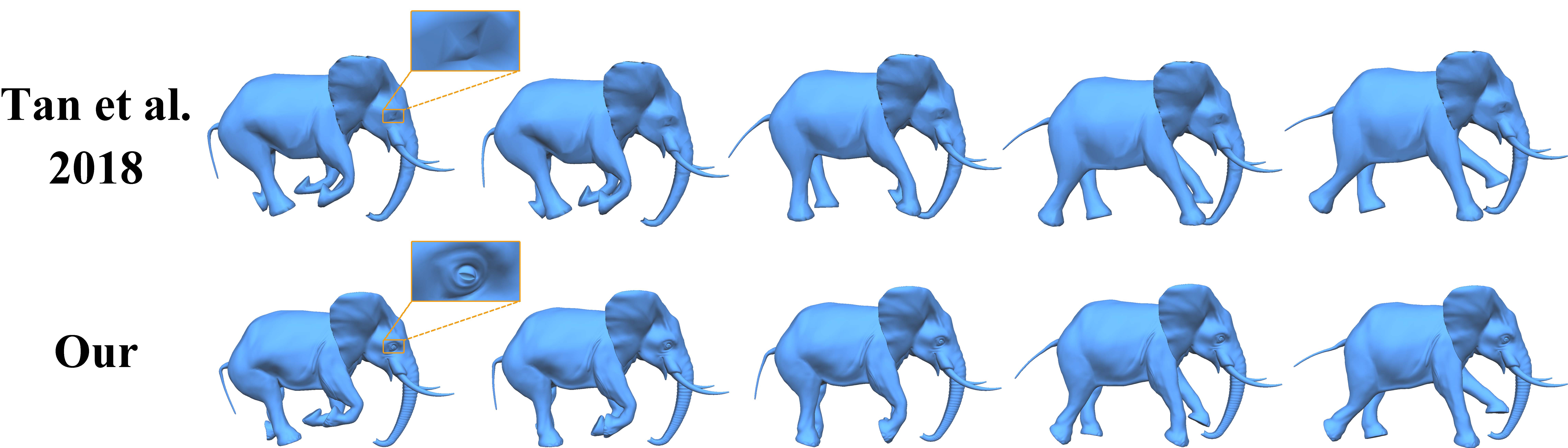}
	\end{center}
	\vspace{-2mm}
	\caption{Interpolation comparison between Mesh VAE~\cite{meshVAE} and our method. The original elephant model~\cite{sumner2004deformation} has 42,321 vertices, which cannot be handled by Mesh VAE due to memory restriction and therefore a simplified mesh with 5,394 vertices is used instead. Our method operates on the original mesh model and produces results with more details.}
	\label{fig:dense_comp}
	\vspace{-3mm}
\end{figure}

\subsection{Mesh Interpolation}
\begin{figure}
	%\vspace{-3mm}
	\begin{center}
		\includegraphics[width=.9\linewidth]{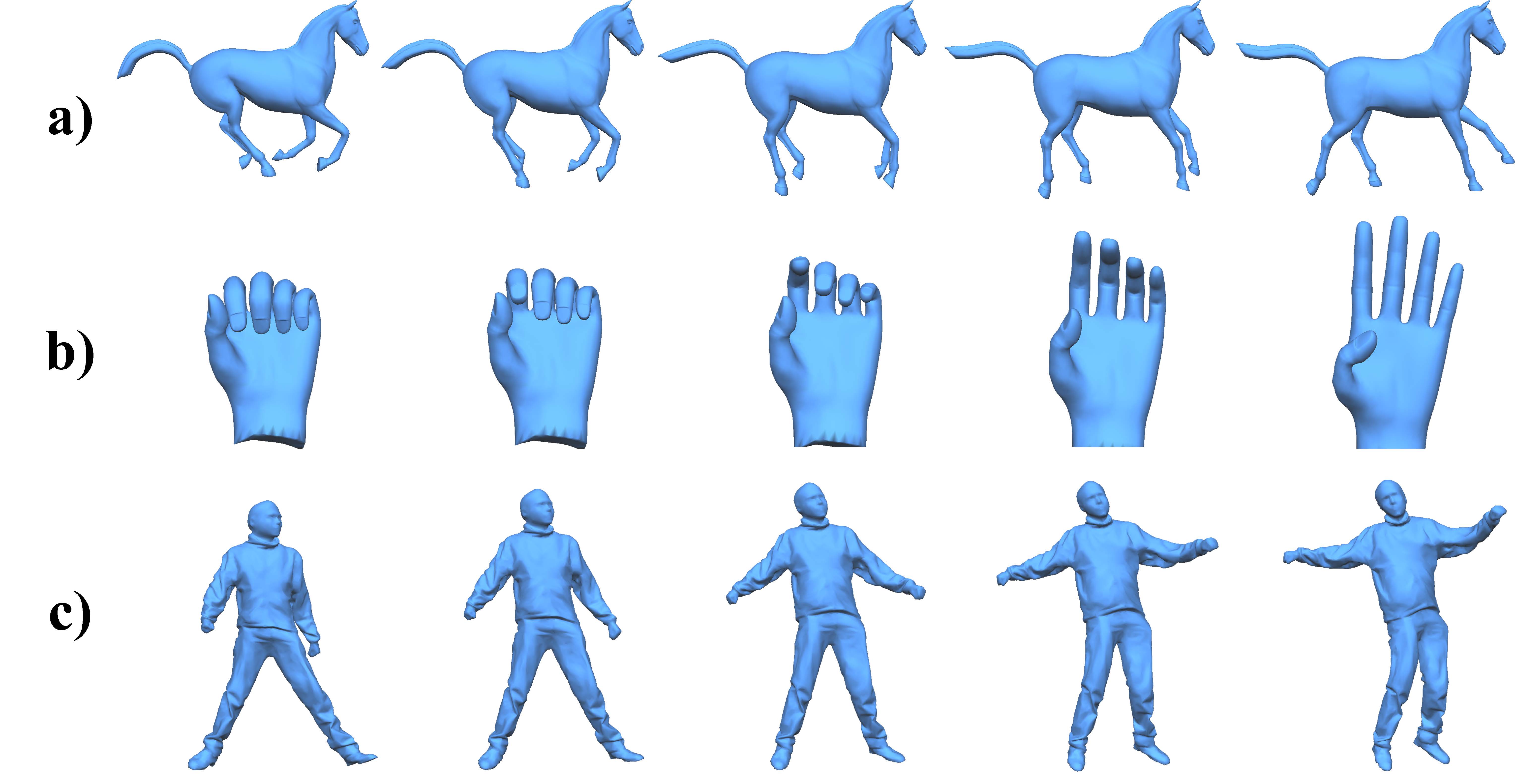}
	\end{center}
	\vspace{-4mm}
	\caption{More interpolation results. (a)(b) more diverse shapes other than human bodies. (c) results interpolated between newly generated shapes.}
	\label{fig:interpolation}
	\vspace{-3mm}
\end{figure}

%For model generation, we have a further application, that is, model interpolation. 
Our method can also be used for shape interpolation. This is because
VAE compresses complex 3D models into a low-dimensional latent vector.
%it can handle details more comprehensively than interpolation directly on the shape. 
We test the capability of our framework to interpolate two different shapes in the original dataset. First, we generate the mean outputs of the probabilistic encoder for two shapes. Then, we linearly interpolate between the two latent vectors and generate a sequence of latent vectors for the probabilistic decoder. Finally, we use the outputs of the probabilistic decoder to reconstruct a 3D deformation sequence. We compare our method on the SCAPE dataset~\cite{anguelov2005scape} with Mesh VAE~\cite{meshVAE}, and a state-of-the-art data-driven deformation method~\cite{CGF:CGF12991}, as shown in Figures~\ref{fig:meshvae_comp} and~\ref{fig:DD_comp} respectively. We can see that Mesh VAE~\cite{meshVAE} produces interpolation results with obvious artifacts. The results by the data-driven method of~\cite{CGF:CGF12991} tend to follow the movement sequences from the original dataset which has similar start and end states, leading to redundant motions such as the swing of right arm. In contrast, our interpolation results give more reasonable motion sequences. {In Fig.~\ref{fig:litany_comp}, we show a comparison with the method of \cite{Litany_2018_CVPR}, which leads to artifacts especially in the synthesized human hands.} 
We show more interpolation results in Fig.~\ref{fig:interpolation}, including sequences between newly generated models and models beyond human bodies.

To show the ability of our network for processing denser meshes, we compare our network with Mesh VAE~\cite{meshVAE}. Mesh VAE can handle much smaller-resolution meshes due to the high memory demands of their fully connected network. This is not a problem with our method thanks to the convolution kernels and pooling operations. Therefore our method can recover superior details for interpolation, reconstruction and random generation. A comparison example for interpolation is shown in Fig.~\ref{fig:dense_comp}.

% , we can conclude that our mesh VAE++ restored model is superior to the mesh VAE model in detail. In fact, not only for interpolation, but also for reconstruction and random generation, mesh VAE++ can produce detailed shapes.

% \begin{figure}
% 	\begin{center}
% 		\subfigure[Dataset Visualization] { \includegraphics[width=0.35\linewidth]{image/ball64.pdf} }
% 		\subfigure[Ball Embedding] { \includegraphics[width=0.55\linewidth]{image/ball_emb1.pdf} }
% 	\end{center}
% 	\vspace{-2mm}
% 	\caption{2D embedding of Ball dataset. (a) the ball dataset and (b) 2D embedding.}\vspace{-3mm}
% 	\label{fig:ball_emb}
% \end{figure}

\subsection{Embedding}
\begin{figure}
	%\vspace{-5mm}
	\begin{center}
		\includegraphics[width=.9\linewidth]{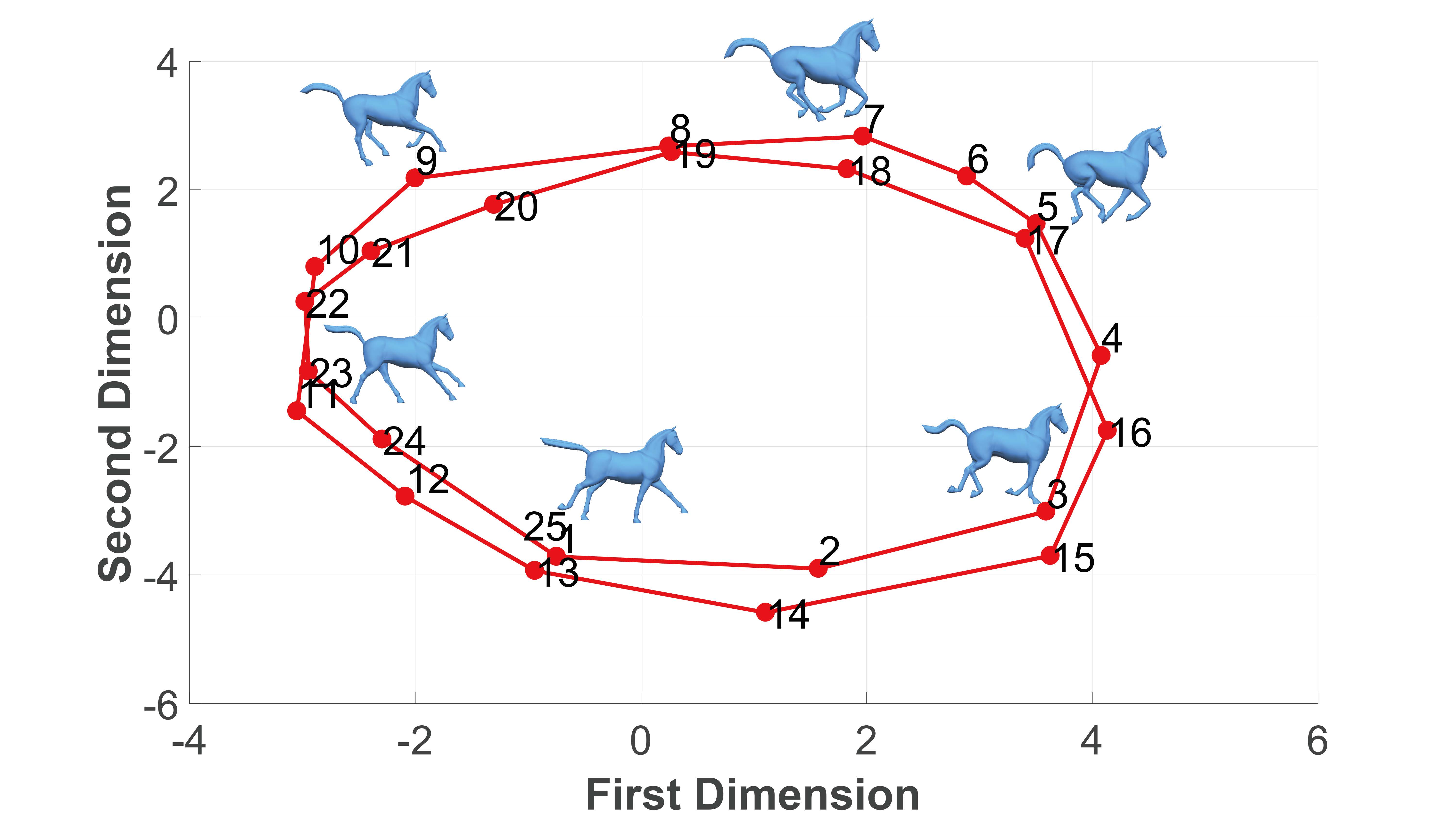}
	\end{center}
	\vspace{-2mm}
	\caption{2D embedding of Horse dataset~\cite{sumner2004deformation}. The result presented as a circle matching the cyclic motion sequence.}\vspace{-3mm}
	\label{fig:horse_emb}
	\vspace{-2mm}
\end{figure}

\begin{figure}
	\centering
	\subfigure[Our]{\includegraphics[width=0.3\linewidth]{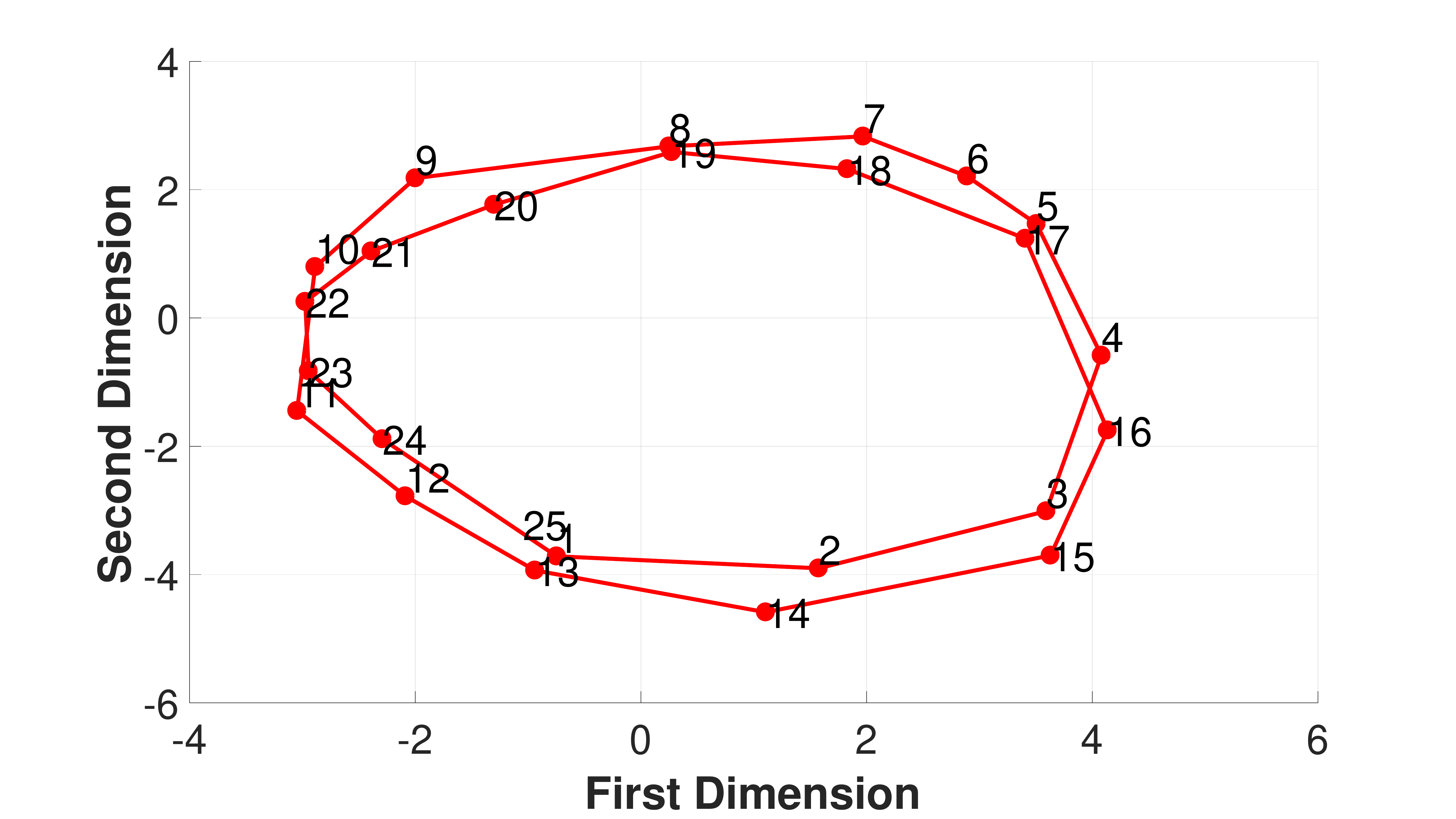}}
	\subfigure[t-SNE]{\includegraphics[width=0.3\linewidth]{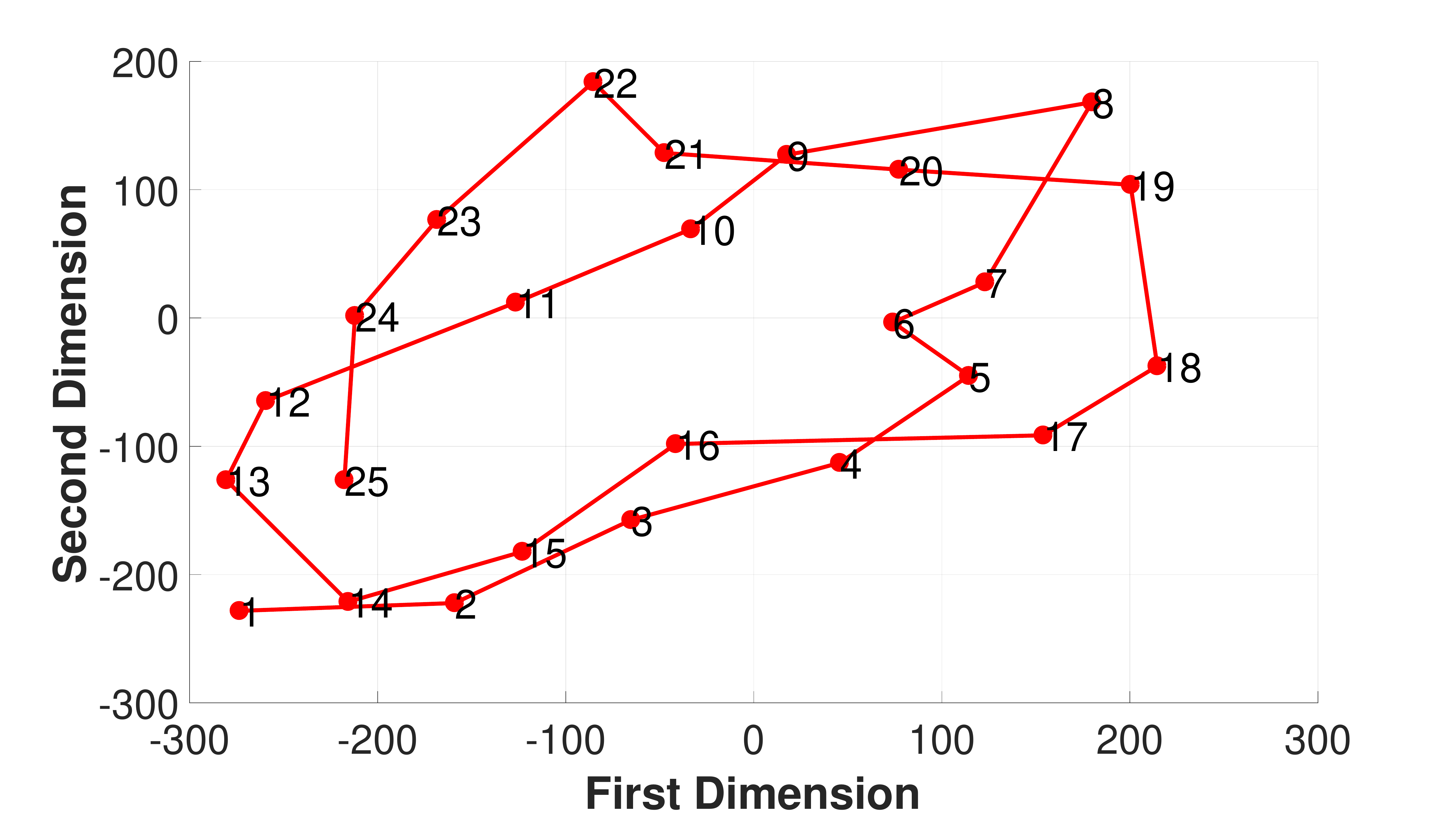}}
	\subfigure[PCA]{\includegraphics[width=0.3\linewidth]{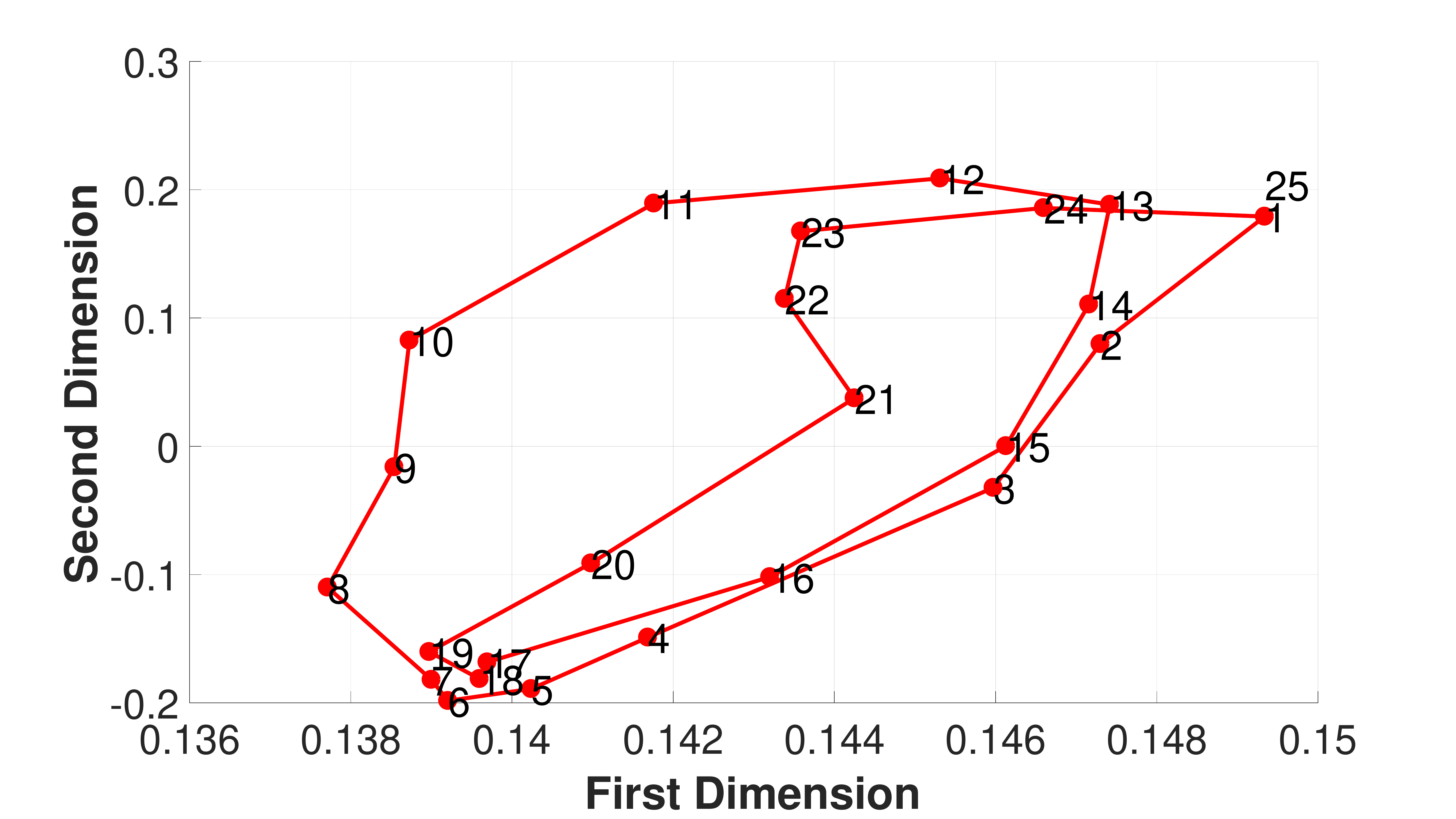}}\vspace{-1mm}
	\caption{Embedding comparisons with t-SNE and PCA.}
	\vspace{-4mm}
	\label{fig:emb_comp}
\end{figure}

Our method can compress 3D shapes into low dimensional vectors for visualization. To better visualize the embedding, we calculate the two largest variances of the latent vector as the horizontal and vertical coordinates of the model in the 2D embedding graph. We utilize this capability to embed shapes in a low-dimensional space.
Our method divides all the models according to their shapes, while allowing models of similar poses to stay in close places. 
We use a representative motion sequences of different deformation types, namely a horse motion sequence from~\cite{sumner2004deformation}.
The Horse dataset~\cite{sumner2004deformation} contains a motion sequence of a galloping horse, which forms a cyclic sequence. We can conclude from the embedding result shown in Fig.~\ref{fig:horse_emb}, that a circle is formed that matches the original sequence, which shows that our network has good embedding ability. \yyj{We also show comparison with t-SNE and PCA in Fig.~\ref{fig:emb_comp}. Our result presents as two circles, while t-SNE and PCA cannot reveal the intrinsic information of the data.}
% Our method effectively divides all the models according to their shapes, while allowing models of similar poses to stay in close places. We use two representative motion sequences of different deformation types, namely a ball deformation sequence slowly bulging from two different directions, and a horse motion sequence from~\cite{sumner2004deformation}. Fig.~\ref{fig:ball_emb} shows the results of ball bulging, where the embedding clearly delineates the deformation behavior.
% The Horse dataset~\cite{sumner2004deformation} contains a motion sequence of a galloping horse, which forms a cyclic sequence. We can conclude from the embedding result shown in Fig.~\ref{fig:horse_emb}, that a circle is formed that matches the original sequence. These two cases show that our network has good embedding ability.

%\vspace{-2mm}
\section{Conclusions}
In this paper we introduced a newly defined pooling operation based on a modified mesh simplification algorithm and integrated it into a mesh variational auto-encoder architecture, which uses per-vertex feature representations as inputs, and utilizes graph convolutions. Through extensive experiments we demonstrated that our generative model has better generalization ability. Compared to the original Mesh VAE, our method can generate high quality deformable models with richer details. Our experiments also show that our method outperforms the state-of-the-art methods in various applications including shape generation and shape interpolation. One of the limitations of our method is that it can process only homogeneous meshes. As a future work, it is desirable to develop a framework capable of handling shapes with different topology as input.

%-------------------------------------------------------------------------

{\small
	\bibliographystyle{cvm}
	\bibliography{cvmbib}
}

\end{document}